\documentclass[12pt, preprint]{aastex}
\usepackage{emulateapj5}
\usepackage{onecolfloat5}
\usepackage{apjfonts}
\usepackage{epsfig}

\def\myputfigure#1#2#3#4#5%
{\vskip#5pt\makebox[0pt]{\hskip#2in
\includegraphics[width=#3\textwidth]{#1}}\vskip#4pt\hfill}

\newcommand\lsim{\mathrel{\rlap{\lower4pt\hbox{\hskip1pt$\sim$}}
        \raise1pt\hbox{$<$}}}
\newcommand\gsim{\mathrel{\rlap{\lower4pt\hbox{\hskip1pt$\sim$}}
        \raise1pt\hbox{$>$}}}

\newcommand{\sfrd}{\dot{\rho}_\ast(z)}

\newcommand{\msun}{M_\odot}
\newcommand{\sneff}{\eta_{\rm SN}}
\newcommand{\Tvir}{T_{\rm vir}}
\newcommand{\zre}{z_{\rm re}}
\newcommand{\vcirc}{v_{\rm circ}}
\newcommand{\kmps}{\rm km~s^{-1}}
\newcommand{\texp}{t_{\rm exp}}
\newcommand{\tsurv}{t_{\rm surv}}

\newcommand{\numfield}{n_{\rm field}}
\newcommand{\aventop}{\langle N_{\rm pre} \rangle}
\newcommand{\avenbot}{\langle N_{\rm post} \rangle}
\newcommand{\zline}{z_{\rm H \alpha ~ 4.5 \mu m}}
\newcommand{\deltazre}{\Delta\zre}
\newcommand{\fSN}{f_{\rm SN}(>t_{\rm obs}, z)}

\def\gsim{\;\rlap{\lower 2.5pt
 \hbox{$\sim$}}\raise 1.5pt\hbox{$>$}\;}
\def\lsim{\;\rlap{\lower 2.5pt
   \hbox{$\sim$}}\raise 1.5pt\hbox{$<$}\;}
 
\begin{document}
\twocolumn[
\title{The Redshift Distribution of Distant Supernovae and its Use in Probing Reionization}

\author{Andrei Mesinger, Benjamin D. Johnson \& Zolt\'{a}n Haiman}
\affil{Department of Astronomy, Columbia University, 550 West 120th Street, New York, NY 10027}
\vspace{+0.4cm}

\submitted{Accepted for publication in the ApJ}

\begin{abstract}
We model the number of detectable supernovae (SNe) as a function of
redshift at different flux thresholds, making use of the observed
properties of local SNe, such as their lightcurves, fiducial spectra,
and peak magnitude distributions. We assume that the star-formation
rate (SFR) at high redshift traces the formation rate of dark matter
halos.  We obtain a rate of $0.4-2.3$ SNe/arcmin$^2$/yr at $z\gsim5$
at the near infrared (4.5~\micron) flux density threshold of 3 nJy
(achievable with the {\it James Webb Space Telescope (JWST)} in a
$10^5$sec integration).  In a hypothetical one--year survey, it should
be possible to detect up to several thousand SNe per unit redshift at
$z\sim6$.  We discuss the possible application of such a large sample
of distant SNe as a probe of the epoch of reionization. By heating the
intergalactic medium (IGM) and raising the cosmological Jeans mass,
the process of reionization can suppress star formation in low--mass
galaxies.  This could have produced a relatively sharp drop in the SNR
around the redshift of reionization ($\zre$).  We quantify the
detectability of this feature in future surveys of distant SNe by
varying the redshift and duration of the feature, as well as its
impact on the SFR in low--mass halos, which results in different
redshifts, widths, and sizes of the drop in the expected SFR.  We find
that the drop can be detected out to $\zre \sim 13$, as long as (i)
the reionization history contains a relatively rapid feature which is
synchronized over different regions to within $\Delta z \lsim$ 1 -- 3,
(ii) the star--formation efficiency in halos that dominate
reionization is $\epsilon_\ast\sim10\%$, and (iii) reionization
significantly suppresses the star formation in low--mass halos.
Depending on the details of (i) -- (iii), this could be achieved with
a survey lasting less than two weeks.  Detecting this signature would
also help elucidate the feedback mechanism that regulates
reionization.
\end{abstract}
\keywords{cosmology: theory -- early Universe -- supernovae: general
-- galaxies: high-redshift -- evolution}
\vspace{+0.5cm}
]

\section{Introduction}
\label{sec:intro}

The formation of the first astrophysical objects and the subsequent epoch of reionization offer a wealth of insight into the physical processes of the early universe.  Two recent developments, revealing the impact of
reionization on the Lyman $\alpha$ absorption spectra of $z\gsim 6$
quasars~\citep{Fan04, MH04, WL04_nf}, and on the
temperature/polarization cross-correlation anisotropy in the cosmic
microwave background (CMB) radiation~\citep{Kogut03, Spergel03},
suggest that reionization history may have been very complex (see, e.g., Haiman 2003 for a review).    Various tools designed to probe this epoch have been proposed, relying on upcoming 21-cm surveys (e.g. \citealt{MMR97}), high-redshift quasar and galaxy spectra (e.g. \citealt{MHC04}), high-redshift galaxy surveys (e.g. \citealt{BL00}), GRBs (e.g. \citealt{BL04_grbvsqso}), and more.  Nevertheless, there is currently a striking dearth of data or theoretical consensus concerning this important period in the universe's evolution.  Due to the considerable technical challenges associated with such high--redshift observations, some of the most useful objects in the near future are probably going to be some of the brightest and hence easiest to detect, such as bright quasars, GRBs and SNe.  Here we make use of the observed properties of local SNe to construct detailed high-redshift SNe rates, obtainable from a future infrared
space telescope such as {\it JWST}. We also show
that such a sample of high--redshift SNe could be an invaluable tool in
probing the reionization epoch, and the physical feedback mechanism
that regulates it.

There have been several previous studies of the expected early SNR.
However, these either did not focus on the expected rates at redshifts
beyond reionization (e.g \citealt{MVP98}), focused exclusively on very
high-$z$ rates from the first generation of metal--free stars
\citep{WA05}, and/or did not address the impact of reionization on the
SFR (e.g. \citealt{MeR97, DF99}).  Likewise, previous theoretical
predictions of the high-redshift SFR \citep{BL00, BL02, CS02} assumed
a fixed degree of suppression due to reionization, matching numerical
simulations which did not include self-shielding \citep{TW96}.  The
main distinctions of the present paper are that we compute the
expected SNe rates for the wide range of redshifts within which
reionization is expected to have occurred ($6 \lsim \zre \lsim 20$), and we
allow various degrees of photo-heating feedback. We also supplement the
standard halo--based estimates of the SFR with a more elaborate
estimate of the corresponding observable SNR, utilizing the observed
properties of low--$z$ SNe.

In the presence of an ionizing background radiation, the IGM is
photo--heated to a temperature of $\gsim 10,000$K, raising the
cosmological Jeans mass, which could suppress gas accretion onto
small-mass halos (e.g. \citealt{Efstathiou92, TW96, Gnedin00b, SGB94}).
Reionization is then expected to be accompanied by a drop in the
global SFR, corresponding to a suppression of star formation in small
halos (i.e. those with virial temperatures below $T_{\rm vir} \lsim$
$10^4$ -- $10^5$ K).  The size of the drop is uncertain, since the
ability of halos to self--shield against the ionizing radiation is
poorly constrained at high redshifts.  Early work on this subject
\citep{TW96} suggested that an ionizing background would completely
suppress star formation in low--redshift ``dwarf galaxy'' halos with
circular velocities $v_{\rm circ} \lsim~35~\kmps$, and partially
suppress star--formation in halos with 35 $\kmps$ $\lsim$ $v_{\rm
circ}$ $\lsim$ 100 $\kmps$.  However, more recent studies \citep{KI00,
Dijkstra04} find that at high--redshifts ($z \gsim 3$), self-shielding
and increased cooling efficiency could be strong countering effects.
These calculations, however, assume spherically symmetry, leaving open
the possibility of strong feedback for a halo with non--isotropic gas
profile, illuminated along a low--column density line of sight (see
\citealt{SIR04} for a detailed treatment of three--dimensional gas
dynamics in photo--heated low--mass halos).

Such a drop in the SFR could be detected in the high--redshift
extension of the 'Lilly -- Madau' diagram \citep{Lilly96, Madau96}, by
directly counting faint galaxies~\citep{BL00}.  In practice, the
low--mass galaxies susceptible to the reionization suppression are
faint and suffer from surface brightness dimming.  Whether or not
these galaxies will fall above the detection threshold of a deep
future survey, so that they can reveal the effect of reionization,
depends crucially on the redshift of reionization, the size of the
affected galaxies and their typical star--formation efficiencies, as
well as the amount of dust obscuration.  Alternatively, by analyzing
the Lyman $\alpha$ absorption spectra of SDSS quasars, \citet{CM02}
suggest that we may already have detected a drop in the SFR at
$z\sim6$, through the non-monotonic evolution of the mean IGM opacity.
In order to improve on this current, low signal--to--noise result,
deep, high-resolution spectra of bright quasars would be required from
beyond the epoch of reionization at $z\gsim6$ (this may be possible
with {\it JWST}, see \citealt{HL99}).

Other events that may trace out the early SFR, such as GRBs and SNe, could be better suited to detect
the reionization drop, as they are bright, and could be used as
tracers of the SFR in arbitrarily faint host galaxies, out to
redshifts higher than the host galaxies themselves.  This is
especially useful, since in hierarchical structure formation
scenarios, the characteristic collapse scale decreases with increasing
redshift. As a result, the individual galaxies that dominate the
ionizing background will likely be undetectable if reionization
occurred at high redshift ($z\gsim 10$). Unfortunately, while GRBs are
bright, they are rare.  Based on the expected {\it Swift} GRB
detection rates (e.g. Figure 6 in \citealt{MPH05}), a drop in the GRB
rate associated with reionization could be detected only at very low,
$\sim1\sigma$, significance, and only if reionization occurred at
$\zre \lsim$ 7 -- 10.  On the other hand, SNe have the benefits of being
both very bright (compared to galaxies at the very faint end of the
luminosity function), and occurring much more frequently (compared to
GRBs).

  In summary, {\it the purpose of this paper is twofold: (i) to construct expected detection rates of high-$z$ SNe in SNe surveys which could be carried out in the future (e.g. with {\it JWST.}); (ii) to quantify the prospects of detecting a reionization feature from such high-redshift SN rates}.

The rest of this paper is organized as follows.  In
\S~\ref{sec:method}, we present our method for predicting the global
SFR and SNR, and estimating the fraction of SNe detectable in a future
survey. In \S~\ref{sec:SNe_detect}, we present our estimates for the SNe detection rates.  In \S~\ref{sec:reion_detect}, we estimate the significance at which the reionization--drop in the SNR may be
detectable.  Finally, we summarize the implications of this work and
offer our conclusions in \S~\ref{sec:conclusions}.

Throughout this paper, we assume standard concordance cosmological
parameters, ($\Omega_\Lambda$, $\Omega_{\rm M}$, $\Omega_b$, n,
$\sigma_8$, $H_0$) = (0.73, 0.27, 0.044, 1, 0.9, 71 km s$^{-1}$
Mpc$^{-1}$), consistent with the recent measurements of the power
spectrum of CMB temperature anisotropies by the {\it Wilkinson
Microwave Anisotropy Probe (WMAP)} satellite \citep{Spergel03}.

\section{Modeling Methods}
\label{sec:method}

\subsection{The Intrinsic Star Formation and Supernova Rates}
\label{sec:SFR_SNint}

Here we briefly outline our method for obtaining the SFR and the
intrinsic SNR, using extended Press-Schechter formalism (EPS) (see
\citealt{LC93}).  The procedure we use to obtain the SFR is based on a
relatively standard approach (see, e.g., \citealt{MPH05} and
references therein).  Readers interested in more details are
encouraged to consult \citet{MPH05}.

We estimate the global SFR density at redshift $z$ as
\begin{equation}
\label{eq:sfrd_general}
\sfrd = \epsilon_\ast \frac{\Omega_b}{\Omega_{\rm M}} \int_{M_{\rm min}(z)}^{\infty} dM \int_{\infty}^{z} dz' M \frac{\partial^2 n(>M, z')}{\partial M \partial z'} P(\tau) ~ ,
\end{equation}
where $\epsilon_\ast$ is the efficiency (by mass) for the conversion
of gas into stars, $M dM (\partial n(>M, z) / \partial M)$ is the mass
density contributed by halos with total (dark matter + baryonic)
masses between $M$ and $M + dM$, $t(z)$ is the age of the universe at
redshift $z$, and $P(\tau)$ is the probability per unit time that new
stars form in a mass element of age $\tau \equiv t(z) - t(z')$
(normalized to $\int_0^\infty d\tau P(\tau)=1$). We adopt the fiducial
value of $\epsilon_\ast = 0.1$ (see, e.g., \citealt{Cen03}). Note that
the star formation efficiency in minihalos (i.e. halos with virial
temperatures below $10^4$K) that contain pristine metal--free gas
could be significantly lower than 0.1, as suggested by numerical
simulations of the first generation of stars \citep{ABN02, BCL02}. The
expected pre--reionization SNe rates would then lie closer to our
$T_{\rm vir} \gsim 10^4$ K curves prior to reionization (see
discussion below), making the detection of the reionization feature
significantly more difficult.  However, the efficiency is likely to be
very sensitive to even trace amounts of metalicity~\citep{BFCL01}, and
conditions for star--formation may result in a standard initial mass
function (IMF) in gas that has been enriched to metaliticites above a
fraction $10^{-4}$ of the solar value.  Indeed, it is unlikely that
metal--free star--formation in minihalos can produce enough ionizing
photons to dominate the ionizing background at reionization
(e.g. \citealt{HAR00, HH03, SYAHS04}).  We assume further that
star--formation occurs on an extended time-scale, corresponding to the
dynamical time, $t_{\rm dyn} \sim [ G \rho(z) ]^{-1/2}$ \citep{CO92,
Gnedin96}:
\begin{equation}
P(\tau) = \frac{\tau}{t_{\rm dyn}^2} \exp \left[ -\frac{\tau}{t_{\rm dyn}} \right] ~ ,
\end{equation}
where $\rho(z) \approx \Delta_c \rho_{\rm crit}(z)$ is the mean mass
density interior to collapsed spherical halos (e.g. \citealt{BL01}),
and $\Delta_c$ is obtained from the fitting formula in \citet{BN98},
with $\Delta_c$ = $18\pi^2$ $\approx$ 178 in the Einstein--de Sitter
model.  The minimum mass, $M_{\rm min}(z)$ in eq. (1), depends on the efficiency
with which gas can cool and collapse into a dark matter halo.  Prior
to reionization and without molecular hydrogen, $M_{\rm min}(z)$
corresponds to a halo with virial temperature, $T_{\rm vir} \sim 10^4$
K; with a significant $\rm H_2$ abundance, the threshold decreases to
$T_{\rm vir} \sim 300$ K (\citealt{HAR00}; we use the conversion
between halo mass and virial temperature as given in \citealt{BL01}).
Post reionization, the Jeans mass is raised, so $M_{\rm min}(z)$ could
increase.  The degree of self-shielding, the ability of the halo gas
to cool, as well as the amount of $\rm H_2$ present in the
high--redshift low--mass halos is uncertain~\citep{Dijkstra04}, and so
below we present results for several values of $M_{\rm min}(z)$, which
we will henceforth express in terms of $T_{\rm vir}$(z).

Next, from $\sfrd$ we obtain the intrinsic differential SNR (number of
core collapse SNe per unit redshift per year) with
\begin{equation}
\label{eq:SNint}
\frac{d\dot{N}}{dz} = \sneff \frac{1}{1+z} \frac{dV(z)}{dz} \sfrd ~ ,
\end{equation}
where the factor $1/(1+z)$ accounts for time dilation, $dV(z)/dz$ is
the comoving volume in our past light cone per unit redshift, and
$\sneff$ is the number of SNe per solar mass in stars.  For a fiducial
Salpeter initial mass function (IMF), we obtain $\sneff \sim
1/180~{\rm M_\odot^{-1}}$, assuming all stars with masses $9~\msun
\lsim M \lsim 40~\msun$ become core collapse SNe \citep{Heger03}. We
neglect the lifetime of these high mass stars in determining our SNR;
this is a reasonable assumption as the lifetimes (as well as the
spread in the lifetimes) are shorter than a unit redshift interval for
redshifts of interest.  Note that an alternative extreme shape for the
IMF, consisting entirely of 100--200 $\msun$ stars \citep{ABN02,
BCL02} would yield a similar value for $\sneff$.

We present our SFR densities ({\it top panel}) and SNRs ({\it bottom
panel}) in Figure~\ref{fig:SFR_SNint}.  The curves correspond to
redshift--independent virial temperature cutoffs of $\Tvir =
300,~10^4,~4.5\times 10^4$, and $1.1 \times 10^5$ K (or circular
velocities of $\vcirc = 3,~17,~35$, and $55~\kmps$, respectively), top
to bottom, spanning the expected range~\citep{TW96, Dijkstra04}.  Also
shown are recent results from GOODS \citep{Giavalisco04}: the bottom
points assume no dust correction and the top points are dust corrected
according to \citep{AS00}; the statistical error bars lie within the
points \citep{AS00}.  As there are large uncertainties associated with
dust correction, each pair of points (top and bottom) serves to
encompass the expected SFR densities.

Our SFRs are consistent with other theoretical predictions
(e.g. \citealt{SPF01,BL00, BL02}), as well as other recent estimates
from the Hubble Ultra Deep Field \citep{Bunker04} and the FORS Deep
Field on the VLT \citep{Gabasch04}, after they incorporate a factor of
5--10 increase in the SFR \citep{AS00} due to dust obscuration.
Furthermore, we note that our SNRs, which at $z\gsim5$ yield 0.3 -- 2
SNe per square arcminute per year, are in good agreement with the
$\sim$ 1 SN per square arcminute per year estimated by \citet{MeR97}
by requiring that high-redshift SNe produce a mean metalicity of $\sim
0.01~Z_\odot$ by $z \sim 5$.  Note, however, that the rates we obtain
are significantly higher (by a factor of $\sim$ 60 -- 2000 at
$z\sim20$) than those recently found by~\citet{WA05}.  The reason for
this large difference is that Wise \& Abel consider star--formation
only in minihalos, and they assume a very low star--formation
efficiency of a single star per minihalo, as may be appropriate for
star--formation out of pristine (metal--free) gas in the first
generation of minihalos~\citep{ABN02, BCL02}. In contrast, we assume
an efficiency of $\epsilon_\ast=0.1$, which may be more appropriate
for star--formation in pre--enriched gas that dominates the SFR just
prior to reionization (including star--formation in minihalos).

As mentioned above, by increasing the cosmological Jeans mass,
reionization is expected to cause a drop in the SFR (and hence the
SNR), with the rates going from the horizontally striped region in
Figure~\ref{fig:SFR_SNint} at $z > \zre$ to the vertically striped
region at $z < \zre$.

The redshift width of this transition is set by a combination of
large-scale cosmic variance, radiative transfer, and feedback effects.
For the majority of the paper, we use $\deltazre \sim 1$ as a rough indicator of the width of the transition we are analyzing.  We distinguish between ``reionization'' and a ``reionization feature'', and use $\deltazre$ as an indicator of the width of the later.  Even with an extended reionization history ($\Delta z \sim 10$), fairly sharp ($\deltazre \lsim 3$) features are likely.   We postpone a more detailed discussion until \S~\ref{sec:sharp}.

The other important factor determining the usefulness of the method
proposed here is the factor by which the SFR drops during the
reionization epoch.  The size of this drop is mediated by the
effectiveness of self-shielding and gas cooling during photo--heating
feedback: i.e. on whether or not the star--formation efficiency is
significantly suppressed in those halos that dominate the SFR and SNR
immediately preceding the reionization epoch.  Given the uncertainties
about this feedback discussed above, we will consider a range of
possibilities below, parameterized by the modulation in the virial
temperature threshold for star--formation during reionization.

\vspace{+0\baselineskip} \myputfigure{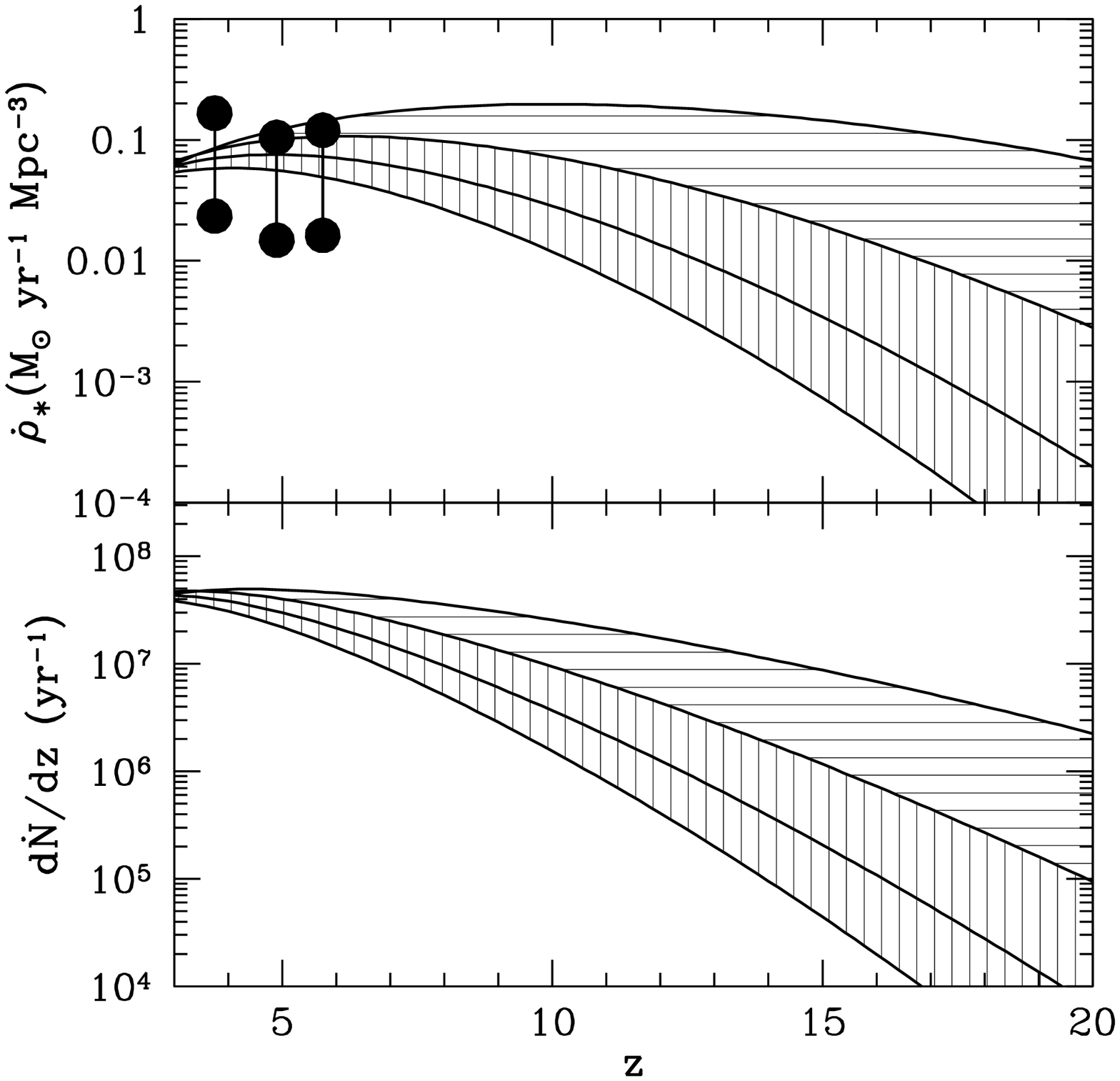}{3.3}{0.5}{.}{0.}
\vspace{-1\baselineskip} \figcaption{ {\it Upper Panel}: SFR densities
obtained from our model.  The curves ({\it top to bottom}) correspond
to different lower cutoffs on the virial temperatures of star--forming
halos, $\Tvir \gsim 300,~10^4,~4.5\times 10^4$, and $1.1 \times 10^5$
K (corresponding to circular velocity thresholds of $\vcirc \gsim
3,~17,~35$, and $55~\kmps$).  Dots indicate results from GOODS
\citep{Giavalisco04}: the lower set of points assume no dust
correction, while the upper set of points are dust corrected;
statistical 1-$\sigma$ error bars lie within the points.  The lines
connecting each pair of points span the expected range of SFR
densities.  {\it Lower Panel}: the SNR densities accompanying the SFR
densities in the top panel.  A reionization feature at redshift $\zre$ is
expected to be evidenced by a drop in the above rates, with rates
changing from the horizontally striped region at $z > \zre$ to the
vertically striped region at $z < \zre$, within a transition period lasting for $\deltazre$.
\label{fig:SFR_SNint}}
\vspace{+1\baselineskip}

\subsection{High Redshift SNe in Future Surveys}
\label{sec:detection}

Given the intrinsic star--formation and SN rates, our next task is to
estimate the number of SNe that could be revealed by a future SNe search.  In general, the number of SNe per unit redshift,
$dN_{\rm exp}/dz$, that are bright enough to be detectable in an
exposure of duration $\texp$ can be expressed as
\begin{equation}
\label{eq:SNexp}
\frac{dN_{\rm exp}}{dz} = \frac{d\dot{N}}{dz} \int_{0}^{\infty} \fSN ~  dt_{\rm obs} ~ ,
\end{equation}
where $(d\dot{N}/dz) dt_{\rm obs}$ is the number of SNe which occurred
between $t_{\rm obs}$ and $t_{\rm obs} + dt_{\rm obs}$ ago (per unit
redshift; note that the global mean SNR will evolve only on the Hubble
expansion time--scale, and can be considered constant over
several years), and $\fSN$ is the fraction of SNe
which remain visible for at least $t_{\rm obs}$ in the observed
frame.  Then the total number of SNe detected in a survey of duration
$\tsurv$ is
\begin{equation}
\label{eq:SNobs}
N_{\rm surv} = \frac{\tsurv}{2~\texp} \frac{\Delta\Omega_{\rm FOV}}{4 \pi} N_{\rm exp} ~ ,
\end{equation}
where $\Delta\Omega_{\rm FOV}$ is the instrument's field of view, and
$\tsurv/(2\texp)$ is the number of fields which can be tiled in the
survey time, $\tsurv$ (we add a factor of 1/2 to allow for a second
pair of filters to aid in the photometric redshift determination; note that this provides for imaging in 4 different {\it JWST} bands; see
discussion below).  Note also that equation (\ref{eq:SNobs}) is
somewhat idealized, in that it assumes continuous integration for the duration of the survey, and e.g., does not account for time required to slew the
instrument to observe different fields. In principle, each field has
to have repeated observations (to detect SNe by their variability),
and therefore any dedicated survey should target fields that have
already been observed. Furthermore, a long, dedicated program may
not be necessary, because the effect could be detected with relatively
few fields (at least under optimistic assumptions; see discussion
below), and several fields with repeated imaging (separated by $>$ 1
-- 2 years) may already be available from other projects; these fields
can then be used for the SN search.

Additionally, it might prove useful to consider the luminosity of a SN host galaxy as a means of biasing the sample towards the small mass halos which are most susceptible to the reionization suppression.  Ignoring SNe which originate in galaxies above a certain threshold luminosity at a given redshift could limit the sample to those small halos which are most affected by reionization, and hence decrease the ``noise'' component in the reionization signal (e.g. eq. (\ref{eq:StoN})).  If there turns out to be
strong correlation between SN type and host halo properties (e.g.
due to metalicity), then, in principle, such correlations can be
empirically discovered and used to further narrow the SN sample
to those most susceptible to feedback - even if the expected 
correlation cannot be predicted ab-initio.

In general, $\fSN$ in equation~(\ref{eq:SNexp}), i.e. the
fraction of SNe which remain visible for at least $t_{\rm obs}$, depends on (i)
the properties of the SN, in particular their peak magnitude and
lightcurve, and the distribution of these properties among SNe, and
(ii) on the properties of the telescope, such as sensitivity, spectral
coverage, and field of view.  In the next two subsections, we discuss
our assumptions and modeling of both of these in turn.

\subsubsection{Empirical Calibration of SN Properties}
\label{sec:empirical}

At each redshift, we run Monte-Carlo simulations to determine $\fSN$ in equation~(\ref{eq:SNexp}). We use the observed properties
of local core--collapse SNe (CCSNe) in estimating $\fSN$.
For the high redshifts of interest here, we only consider core collapse SNe of Type II.
SNe resulting from the collapse of Chandreshekar--mass white dwarfs (Type
Ia) are expected to be extremely rare at high redshifts ($z \gsim 6$), as
the delay between the formation of the progenitor and the SN event ($\gsim
1$ Gyr; \citealt{Strolger04}) is longer than the age of the universe at these
redshifts.  Local core CCSNe come in two important varieties, types
IIP and IIL, differentiated by their lightcurve shapes. We ignore the
extremely rare additional CCSN types, e.g Type IIn and IIb, which
appear to have significant interaction with circumstellar material and
constitute less than 10\% of all CCSNe. Type Ib/c, which may or may not also
result from core collapse, have luminosities and light-curves that are
similar to Type IIL and occur less frequently. While the relative numbers
of Type IIP and Type IIL SNe are not known even for nearby SNe, recent
estimates imply that they are approximately equal in frequency
\citep{C97}. We therefore assume that 50\% of the high-redshift SNe
are Type IIP and 50\% are Type IIL.

CCSNe result from the collapse of the degenerate cores of high-mass
stars.  The luminosity of CCSNe is derived from the initial shock
caused by the core-collapse which ionizes material and fuses unstable
metal isotopes (see \citet{LS03} and references therein for a more
detailed description of SN lightcurves).  In the early stages of the
SN, the shock caused by the core collapse breaks out from the surface
of the progenitor (typically high mass red giants), resulting in a
bright initial peak in the light curve that lasts less than a few days
in the rest--frame of the SN.  As the shock front cools, the SN
dims. However, the SN may then reach a plateau of constant luminosity
in the light curve, believed to be caused by a wave of recombining
material (ionized in the shock) receding through the envelope.  The
duration and strength of this plateau depends on the depth and mass of
the progenitor envelope, as well as the explosion energy, with those SNe
exhibiting a strong plateau classified as Type IIP. A typical plateau
duration is $\lsim 100$ rest--frame days \citep{P94}. In Type IIL SNe,
this plateau is nearly non-existent, and the lightcurve smoothly
transitions from the rapid decline of the cooling shock to a slower
decline where the luminosity is powered by the radioactive decay of
metals in the SN nebula.  After the plateau, Type IIP SNe also enter
this slowly declining `nebular' phase.

These observationally determined behaviors have been summarized in a
useful form as lightcurve templates in \citet{DB85}. We use these
template lightcurves in determining $\fSN$, and normalize
the lightcurves using gaussian--distributed peak magnitudes
(i.e. log--normally distributed in peak flux) determined by
\citet{Richardson02} from a large sample of local Type IIP and Type
IIL SNe (see Table~\ref{tbl:peaks}). We perform the Monte-Carlo
simulations with both the dust corrected, and dust uncorrected values
in \citet{Richardson02}, since the dust production history of the
early universe is poorly understood and is essentially unconstrained
empirically.

We use a combined high-resolution HST STIS + ground--based spectrum of
the Type IIP supernova SN1999em (the `November 5th' spectrum of
\citealt{Baron00}) as the template SN spectrum in order to obtain
$K$-corrections (with the \citealt{DB85} lightcurves given in the
restframe $B$ filter).  This spectrum was obtained within 10 days of
maximum light, i.e. during the initial decline of the SN brightness,
and has been dereddened by A$_{V}=0.3$ mag \citep{Baron00, H01}.
While the spectrum, and hence the $K$-corrections, of SNe evolve
during the lightcurve, this effect is not strong for the wavelengths
of interest \citep{P94}, especially since the lightcurve template we
use is well matched to the wavelengths being probed by the
observations we consider below (leading to small $K$-corrections).  We
have also used this Type IIP SN template spectrum to calculate
$K$-corrections for Type IIL SNe.  This is necessary due to the lack
of restframe UV spectra of Type IIL SNe that can be combined with
optical spectra, and justifiable because the $K$-corrections are
relatively small and the broadband colors of both Type IIP and IIL SNe
(a measure of the spectral shape that determines the $K$-corrections)
are similar, at least in the optical \citep{P94}.  Note that we assume
that very-high redshift core-collapse SNe are similar to local SNe in
their spectra, peak luminosities, and temporal evolution. However,
these assumption do not significantly impact our conclusions below, as
long as (i) the average properties of the SNe do not change
\emph{rapidly} at high redshift (which could mimic the reionization
drop), (ii) they do not become preferentially underluminous (which
would make high-$z$ SNe less detectable, lowering the statistical
confidence at which the reionization drop is measured), and (iii) the
detection efficiency does not change rapidly with redshift (e.g. due
to instrument parameters or spectral lines).

\begin{table}[ht]
\caption{Means and standard deviations of the adopted peak absolute
magnitudes.  Values are taken from \citet{Richardson02}. Note: an
$M_B=-17$ SN would be detectable out to $z \approx 8.2$ at the flux threshold
of 3 $\rm nJy$ at 4.5 $\mu$$\rm m$. }
\vspace{-0.8cm}
\label{tbl:peaks}
\begin{center}
\begin{tabular}{ccccc}
\tablewidth{3in}\\
\hline
\hline
  & \multicolumn{2}{c}{Corrected for Dust} & \multicolumn{2}{c}{Not Corrected for Dust}\\
\hline
SN Type & $\langle M_{B} \rangle$ & $\sigma$ & $\langle M_{B} \rangle$ & $\sigma$\\
\hline
\hline
IIP & -17.00 & 1.12 & -16.61 & 1.23\\
IIL & -18.03 & 0.9 & -17.80 & 0.88\\
\hline
\hline
\end{tabular}\\
\end{center}
\end{table}

In order to use the method outlined above to probe reionization, the
SN redshifts must also be known to an accuracy of $\Delta z\lsim 1$.
SN redshifts can be determined via spectroscopy of either the SN
itself, or of the host galaxy.  However, as we have already noted, the
host galaxies may only be marginally detectable even in imaging, and
the SNe may be too faint for anything other than extremely low
resolution spectroscopy.  We present here only a very brief example of
the possibility of obtaining redshifts from the extremely low
resolution ($ \lambda /\Delta \lambda \sim 5$) spectra provided by
multiband imaging: a complete investigation of this possibility is
warranted, but is beyond the scope of this paper.

To the extent that Type II SNe spectra can be represented as a
sequence of blackbodies of different temperatures
(e.g. \citealt{DF99}) photometric redshifts will be impossible to
obtain without information about the SN epoch, since temperature and
redshift would be degenerate.  However, local Type IIP SNe show
significant deviations from a blackbody in the UV ($\lambda <
3500$\AA) due to metal-line blanketing in the SN photosphere,
providing spectral signatures that could be used as redshift
indicators, depending on their strength.  In Figure~\ref{fig:photoz},
we show the evolution with redshift of the infrared colors (in bands
accessible with {\it JWST}; see below) of our template SN spectrum,
compared with the color evolution of a blackbody. The figure shows
that the template spectrum deviates significantly from a blackbody. If
the spectrum of the SN is always the same as the template spectrum,
then there are good prospects for obtaining photometric redshifts for
these SNe, at least in the redshift range $z=7-13$.  While the figure
shows that multiple redshifts may be possible at fixed observed
colors, the degenerate solutions would correspond to $z>16$ SNe;
contamination from such high redshift will be mitigated by the fact
that these SNe are likely to be too faint to be detected.  Of course,
more detailed studies of the UV behavior of local SNe, especially
their variety and spectral evolution, will be necessary to confirm the
possible use of photometric redshifts.

\vspace{+0\baselineskip}
\myputfigure{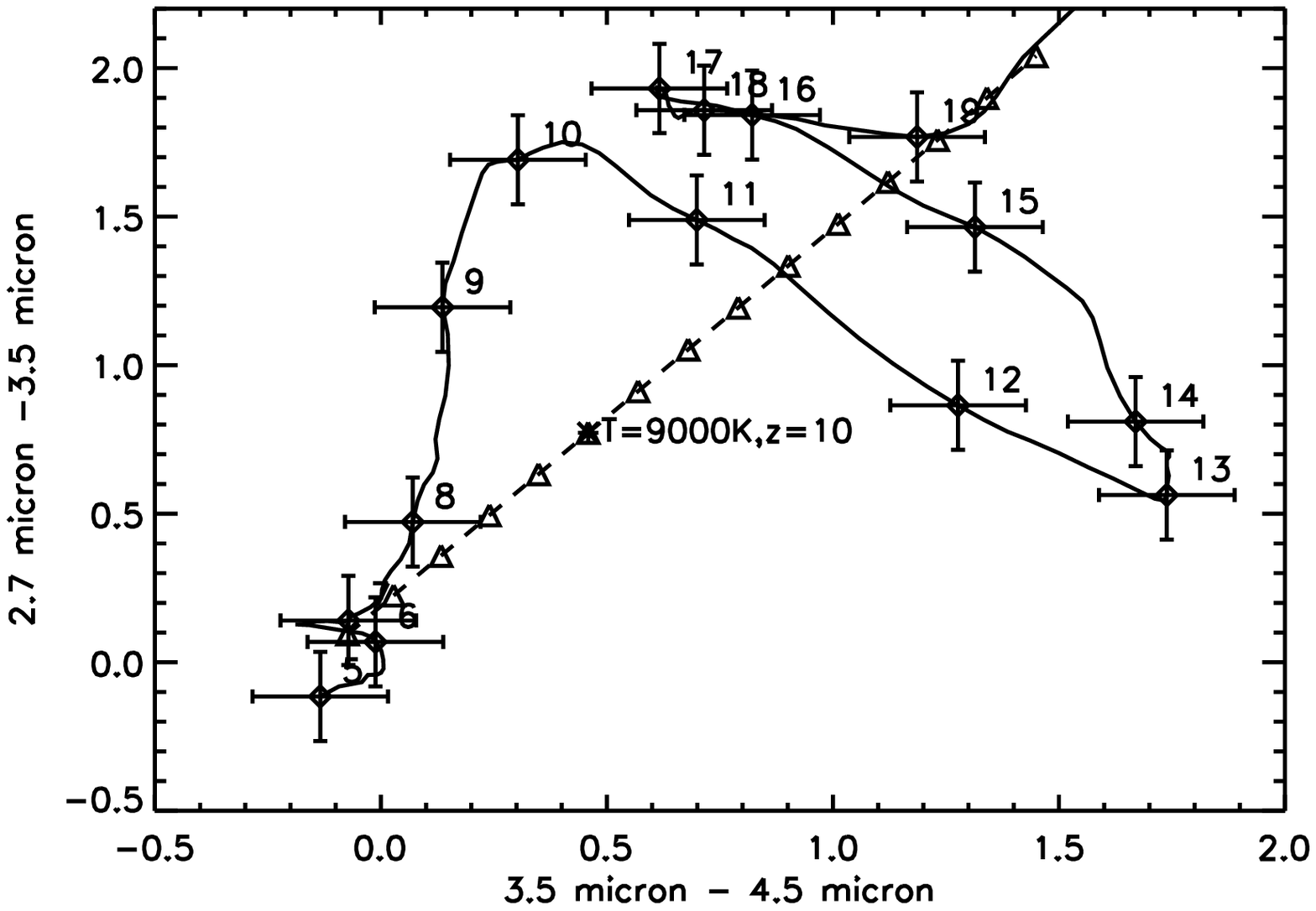}{3.3}{0.5}{.}{0.}
\vspace{-1\baselineskip} \figcaption{Infrared colors of a Type II SN
as a function of its redshift.  \emph{Solid curve}: The template
spectrum of the text, from the Type IIP SN 1999em. Diamonds are placed
at intervals of $\Delta z=1$, labeled with the redshift.  Error bars
of 0.15 mag are shown to give a sense of the photometric errors that
may be expected, excluding the many possible systematic effects.
\emph{Dashed curve:} The colors of a blackbody at different redshifts
(temperatures).  The asterisk marks the color of a 9000K blackbody at
$z=10$, triangles are at intervals of $\Delta z=1$ (or at a constant
redshift but with a corresponding different temperature)
\label{fig:photoz}}
\vspace{+1\baselineskip}

\subsubsection{SNe Detectability and Survey Parameters}
\label{sec:survey}

As a specific example for the number of detectable SNe in a future
sample, we consider observations by {\it JWST}, a 6m diameter space
telescope, scheduled for launch in 2013.  Similar constraints could be delivered by the {\it Joint, Efficient Dark-energy Investigation (JEDI)}\footnote{http://jedi.nhn.ou.edu/} \citep{Wang04}, a proposed instrument which will require a $\sim$ 100 times longer integration time to reach the same detection threshold, but has a $\sim$ 100 times larger FOV.  The relevant instrument on
{\it JWST} is NIRcam,\footnote{See
http://ircamera.as.arizona.edu/nircam for further details.} a
near--infrared imaging detector with a FOV of $2.3'\times4.6'$. A
field can be observed in two filters simultaneously. NIRcam will have
five broadband filters (with resolution $ \lambda / \Delta \lambda
\sim 5$).  We model the filter response as tophat functions with
central wavelengths of 1.5, 2.0, 2.7, 3.5, and 4.5 \micron. For
concreteness, below we will present results only for the 4.5 and 3.5
\micron\ filters, since they are the longest--wavelength {\it JWST}
bands; however, we allow time for imaging in two other bands, if
needed for photometric redshift determinations.  The current estimate
of the {\it JWST} detection threshold at 4.5 \micron~is $\gsim$ 3 nJy
for a 10 $\sigma$ detection and an exposure time of $10^5$ s.  The 3.5
\micron~band is more sensitive, with a detection threshold of $\gsim$
1 nJy for a 10 $\sigma$ detection and an exposure time of $10^5$ s.

We show our results for $\fSN$ in
Figure~\ref{fig:f_hist}.  The solid curves correspond to $z=7$, the
dashed curves correspond to $z=10$, and the dotted curves correspond
to $z=13$.  In each panel, the top set of curves assumes a flux
threshold of 3 nJy (or an exposure time of $t_{\rm exp}=10^5$ seconds
in the 4.5 \micron~band), and the bottom set assumes 9.5 nJy ($t_{\rm
exp}=10^4$ seconds in the 4.5 \micron~band, background dominated).
The top panel further assumes no dust extinction, while in the bottom
panel, we adopt the same dust extinction as in the low redshift sample
\citep{Richardson02}.  Understandably, the distributions get wider as
redshift increases (due to time dilation), but the total visible
fraction, $f_{\rm SN}(>0, z)$, gets smaller (due to the increase in
luminosity distance).  The double bump feature in some of the curves
corresponds to the plateau of Type IIP SNe lightcurves discussed
above.

\vspace{+0\baselineskip}

\myputfigure{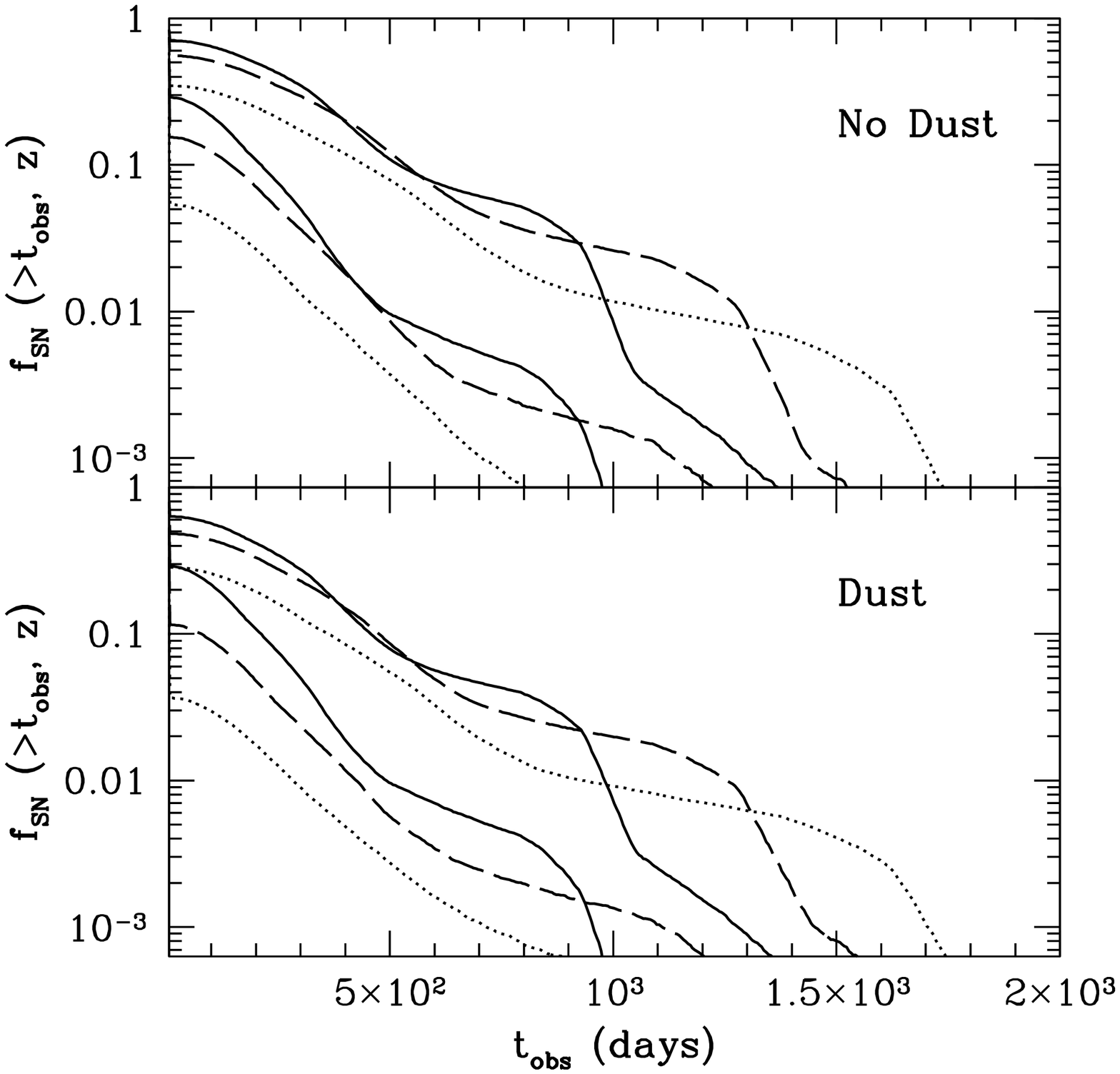}{3.3}{0.5}{.}{0.}
\vspace{-1\baselineskip} \figcaption{ Fraction of SNe which remains
visible for an observed duration of $t_{\rm obs}$ or longer.  The
curves correspond to SN redshifts of $z=7$ ({\it solid curve}), $z=10$
({\it dashed curve}), and $z=13$ ({\it dotted curve}).  In each panel,
the top set of curves assumes a flux threshold of 3 nJy (exposure time
of $t_{\rm exp}=10^5$s with the 4.5 \micron~{\it JWST}), and the bottom set assumes
9.5 nJy ($t_{\rm exp}=10^4$ s).  The top panel assumes no dust
extinction, while the bottom panel assumes the same dust extinction as
observed within the low redshift sample \citep{Richardson02}.
\label{fig:f_hist}}
\vspace{+1\baselineskip}

As seen in Figure \ref{fig:f_hist}, most of the supernovae that are
bright enough to be visible at all, will remain visible for up to
$\sim$ 1 -- 2 years.  Hence, in order to catch the most SNe, it will
be necessary to have repeat observations of the SN survey fields a few
years apart, to insure that most of the observed SNe will be
identified as new sources or sources that have disappeared.  A time
between observations that is comparable to or larger than the SN
duration will be the optimal strategy for detecting the most SNe
\citep{N03}.  Note that we ignore the time required to collect
reference images, since observations conducted for other programs will
likely provide a sufficient set of such reference images.  However, as
shown in equation~(\ref{eq:SNobs}), we allow time for the field to be
imaged in four different bands, to aid in photometric redshift
determination.

To be more explicit, we find that in order to obtain the largest
number of high--redshift SNe, in general it is a more efficient use of
{\it JWST}'s time to 'tile' multiple fields rather than 'stare' for
extended periods of time ($\gsim 10^4$ s) at the same field
\citep{N03}.  This is because of strong time dilation at these
redshifts.  As can be seen from Figure~\ref{fig:f_hist}, most
detectable SNe will remain above the detection threshold for several
months, even assuming $10^4$ s integration times.  Also evident from
Figure~\ref{fig:f_hist} is that the increase in the total visible
fraction of SNe going from $\texp=10^4$ s to $\texp=10^5$ s, is less
than the factor of 10 increase in exposure time. As a result, a
fiducial 1--yr {\it JWST} survey would therefore detect more SNe using
$\texp=10^4$ s than using $\texp=10^5$ s (see
Figures~\ref{fig:SNobs_45}a and \ref{fig:SNobs_35}a).  Understandably,
this conclusion does not hold for very high-redshifts, $z\gsim14$,
where SNe are extremely faint, and require very long exposure times to
be detectable.  However, even with such long exposure times, very few
SNe will be detectable at these large redshifts, rendering the use of
longer exposure times unnecessary.

\section{Results and Discussion}
\label{sec:results}

\subsection{SNe Detection Rates}
\label{sec:SNe_detect}

The number of SNe that could be detectable in putative future surveys
are shown in Figures \ref{fig:SNobs_45}a and \ref{fig:SNobs_35}a.  The
curves correspond to the same virial temperature cutoffs for
star--forming halos as in Figure~\ref{fig:SFR_SNint}.  Solid lines
assume no dust obscuration; dashed lines include a correction for dust
obscuration as discussed above.  Figure \ref{fig:SNobs_45}a shows
results assuming flux density thresholds of 9.5 nJy (or $\texp = 10^4$
s with the 4.5 \micron~{\it JWST} filter) ({\it top panel}) and 3 nJy
($\texp=10^5$ s) ({\it bottom panel}).  Figure \ref{fig:SNobs_35}a
shows results with the 3.5 \micron~filter assuming equivalent exposure
times: flux density thresholds of 3.2 nJy ($\texp = 10^4$ s with the
3.5 \micron~{\it JWST} filter) ({\it top panel}) and 1 nJy
($\texp=10^5$ s) ({\it bottom panel}).  The right vertical axis
displays the number of SNe per unit redshift per FOV ($2.3' \times 4.6'$); the left
vertical axis shows the number of SNe per unit redshift in a fiducial
1--year survey.  As mentioned above, reionization should be marked by
a transition from the region bounded by the top two solid curves to
the region enclosed by the bottom three solid curves (or the analog
with the dashed curves if dust is present at the time of
reionization).

We note that our expected rates are somewhat higher than those in
\citet{DF99}, a previous study which included SNe lightcurves and spectra in the
analysis.  For example, we find 4 -- 24 SNe per field at $z\gsim5$ in
the 4.5 \micron~filter with $\texp=10^5$ s, compared to $\sim 0.7$ SNe
per field at $z\gsim5$ obtained by \citet{DF99} (after updating their
{\it JWST} specifications to the current version).  However, they use
SFRs extrapolated from the low-redshift data available at the time,
which are not a good fit to recent high-$z$ SFR estimates
\citep{Giavalisco04, Gabasch04, Bunker04}, and are lower than our
$z\gsim5$ SFRs by a factor of 6 -- 40.\footnote{The possibility that
the SFR in a ``Lilly-Madau diagram'' remains flat, or even increases,
at redshifts $z\gsim 5$, owing to star--formation in early, low--mass
halos, is also expected theoretically (see, e.g., Fig.1 in
\citealt{BL02} and associated discussion).}  Taking this factor into
account, their procedure yields 4 -- 27 SNe per field at $z\gsim5$,
which is in excellent agreement with our estimate of 4 -- 24 SNe per
field at $z\gsim5$.

\subsection{Detecting Reionization Features}
\label{sec:reion_detect}

To quantify the feasibility of detecting the reionization feature, we
assume a Poisson S/N,
\begin{equation}
\label{eq:StoN}
{\rm S/N} = \sqrt{\numfield} \frac{\aventop - \avenbot}{\sqrt{\aventop
+ \avenbot}} ~ ,
\end{equation}
where $\numfield$ is the number of {\it JWST} fields, and $\aventop$
and $\avenbot$ are the number of observable SNe per field per unit
redshift pre and post reionization, averaged over $\zre < z <
\zre+\deltazre$ and $\zre-\deltazre < z < \zre$, respectively.  In our
fiducial model we take $\deltazre=1$, but below we also investigate
the impacts of a larger value of $\deltazre$.  Equation~(\ref{eq:StoN}) yields a simple figure of merit that does not
distinguish between the theoretical means of the distributions we
generate and the future observational samples which will have to be
interpreted as numbers drawn from distributions with unknown
means. Likewise, simple $\sqrt{N}$ errors are underestimates for small
$N$.  However, we have verified that at our

\twocolumn[
\vspace{+0\baselineskip} 
\epsscale{2.3}
\plottwo{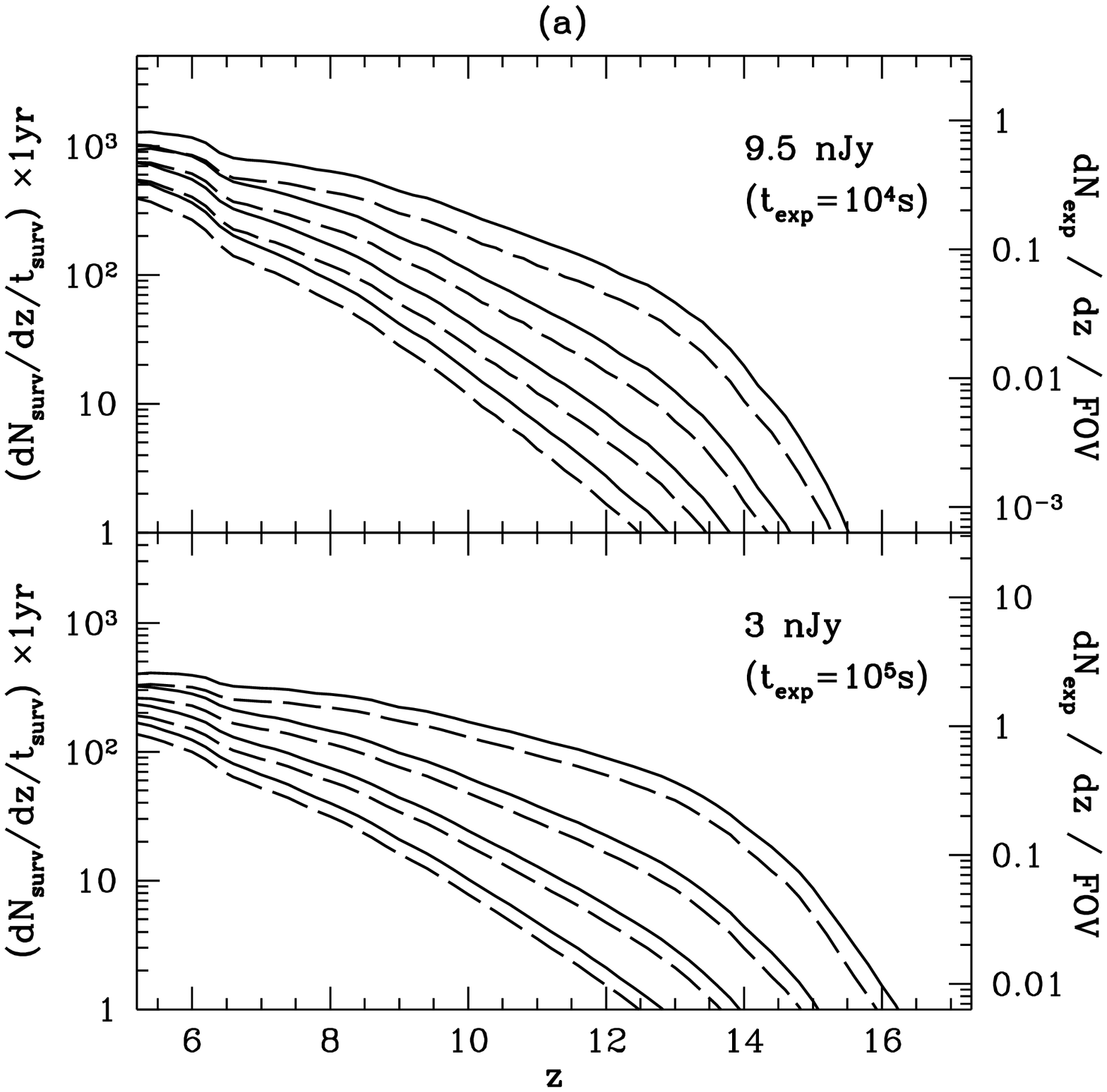}{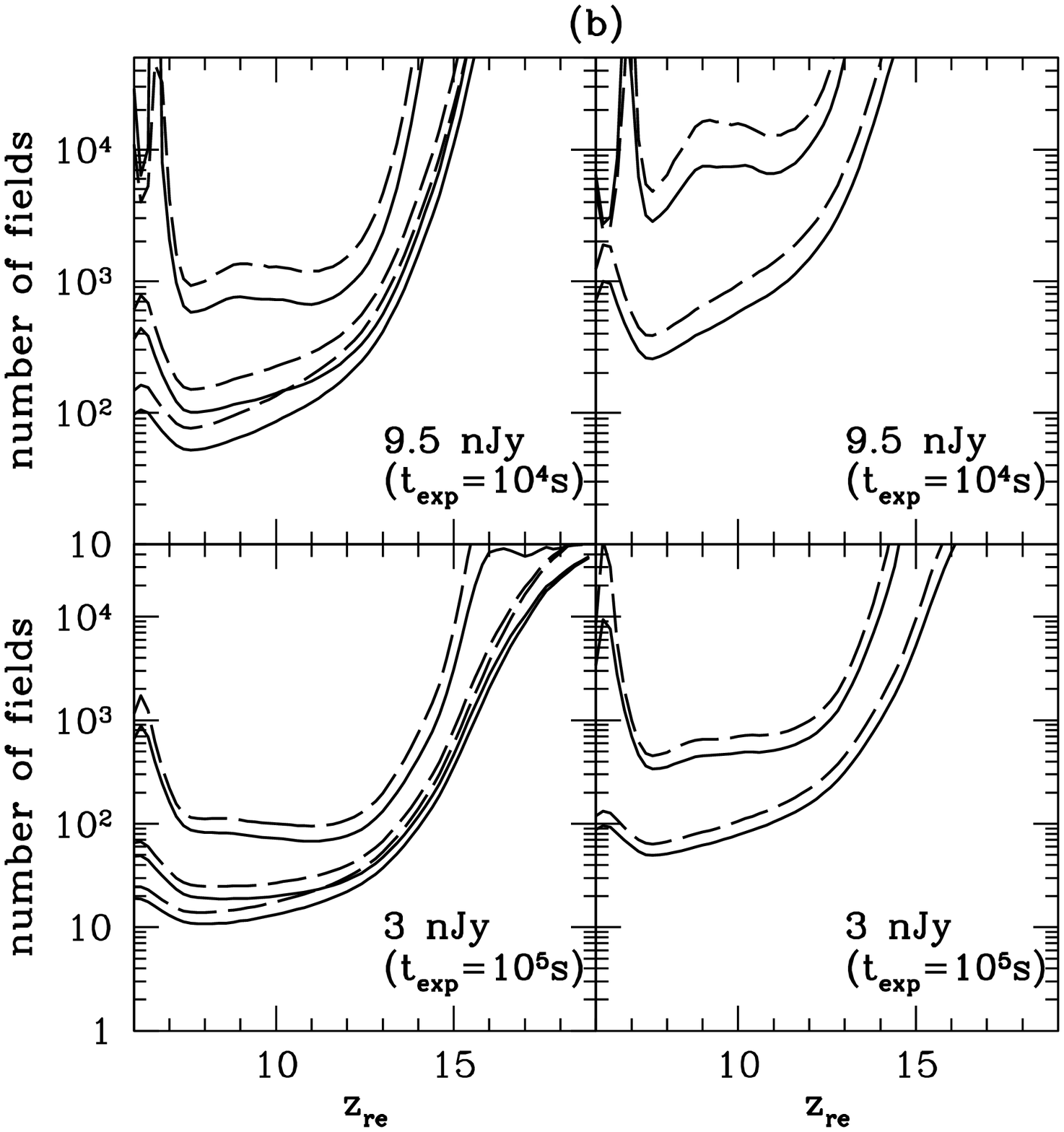}
\figcaption{
(a) Number of high-redshift SNe detectable with the 4.5 \micron~{\it
JWST} filter.  The curves correspond to the same virial temperature
cutoffs as in Figure~\ref{fig:SFR_SNint}.  Solid curves assume no dust
obscuration; dashed curves adopt dust obscuration in the same amount
as observed in the low redshift SNe sample.  The figure shows results
assuming flux density thresholds of 9.5 nJy (or $\texp = 10^4$ s with
{\it JWST}) ({\it top panel}) and 3 nJy (or $\texp=10^5$ s with {\it
JWST}) ({\it bottom panel}).  The right vertical axis displays the
number of SNe per unit redshift per field; the left vertical axis
shows the number of SNe per unit redshift found in $\tsurv/(2\texp)$
such fields (i.e. the differential version of eq. (\ref{eq:SNobs})
with $\tsurv=1$ yr).  Reionization should be marked by a transition
from the region bounded by the top two solid curves to the region
enclosed by the bottom three solid curves (or the analog with the
dashed curves if dust is present at the time of reionization).  (b)
Number of $2.3' \times 4.6'$ {\it JWST} fields required to detect
a reionization feature occurring at $\zre$, with a S/N $\gsim$ 3.  Solid curves
correspond to no dust obscuration; dashed curves include dust.  The
top panels assume $\texp=10^4$ s; bottom panels assume $\texp=10^5$ s.
{\it Left}: Reionization drops in the SNR analogous to $T_{\rm vir}
\gsim 300$ K $\rightarrow$ $10^4$ K, $T_{\rm vir} \gsim 300$ K
$\rightarrow$ $4.5\times10^4$ K, and $T_{\rm vir} \gsim 300$ K
$\rightarrow$ $1.1\times10^5$ K, ({\it top to bottom}).  {\it Right}:
Reionization drops in the SNR analogous to $T_{\rm vir} \gsim 10^4$ K
$\rightarrow$ $4.5\times10^4$ K, and $T_{\rm vir} \gsim 10^4$ K
$\rightarrow$ $1.1\times10^5$ K, ({\it top to bottom}).
\label{fig:SNobs_45}
}
\vspace{+1.5\baselineskip}

\vspace{+0\baselineskip}
\epsscale{5.17}
\plottwo{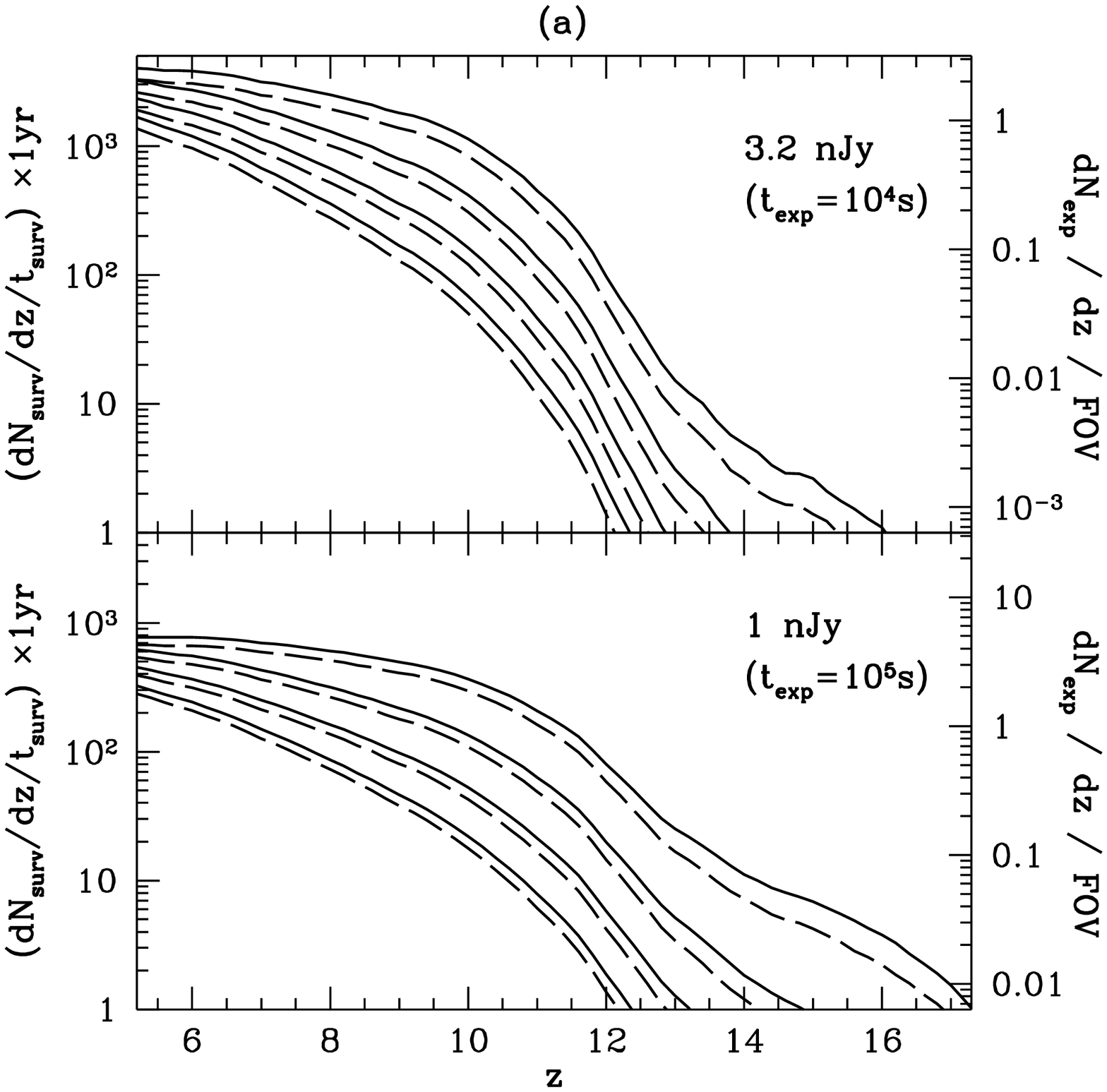}{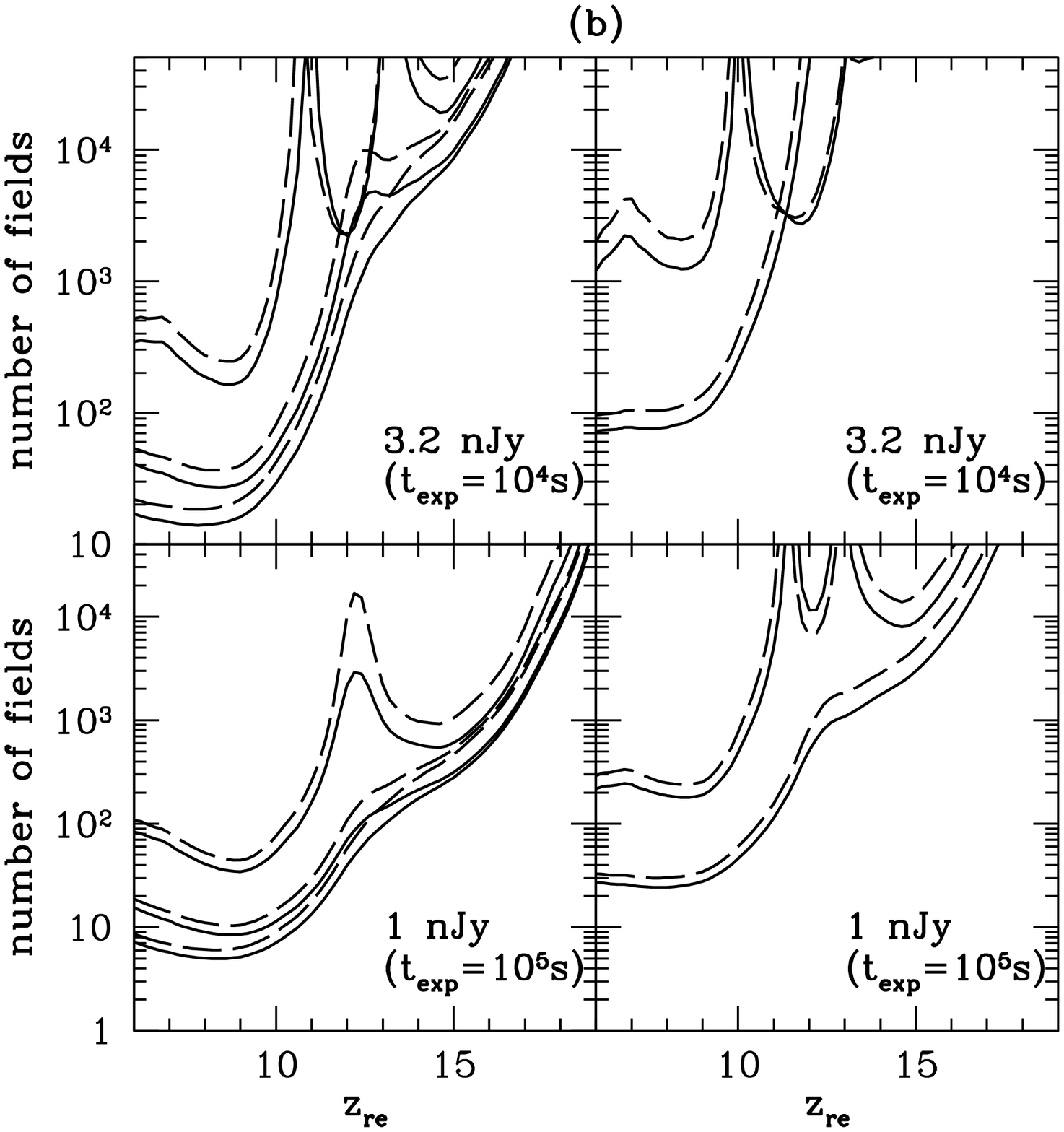}
\vspace{-1\baselineskip} 
\figcaption{Same as Figure~\ref{fig:SNobs_45}, but with the 3.5
\micron~filter, instead of the 4.5 \micron~one.  Note that we present
results for comparable exposure times, hence the sensitivity
thresholds are three times lower than in Figure~\ref{fig:SNobs_45},
due to the disparate sensitivities of the 3.5 \micron~and 4.5
\micron~filters.
\label{fig:SNobs_35}
}
\vspace{+1.5\baselineskip}
]

\noindent
chosen limiting value of
S/N $\gsim$ 3, these issues have little impact on the results.  A full statistical treatment (e.g. comparing redshift distribution of counts
with a Fisher matrix technique; \citealt{HHM01}) is beyond the scope
of this paper, and would be of little value given the uncertainties in
our estimates.  However, we note that the statistics could be improved by combining several (e.g. $\Delta z=0.5$ wide) bins prior to and
following reionization, or by applying the null--hypothesis of a
smoothly evolving SNR directly to the unbinned table of the observed
SNe~\citep{HHM01}.

In Figures \ref{fig:SNobs_45}b and \ref{fig:SNobs_35}b, we plot the
number of 4.5 and 3.5 \micron~{\it JWST} fields, respectively,
required to detect the reionization feature at $\zre$ with a S/N
$\gsim$ 3.  Solid curves correspond to no dust obscuration; dashed
curves include dust corrections.  The top panels assume $\texp=10^4$
s; bottom panels assume $\texp=10^5$ s.  The left panels show the
results assuming that reionization modifies the virial temperature
cutoff as $T_{\rm vir} \gsim 300$ K $\rightarrow$ $10^4$ K, $T_{\rm
vir} \gsim 300$ K $\rightarrow$ $4.5\times10^4$ K, and $T_{\rm vir}
\gsim 300$ K $\rightarrow$ $1.1\times10^5$ K, from top to bottom.  The
right panels show results under the less pronounced transitions of
$T_{\rm vir} \gsim 10^4$ K $\rightarrow$ $4.5\times10^4$ K, and
$T_{\rm vir} \gsim 10^4$ K $\rightarrow$ $1.1\times10^5$ K, from top
to bottom.  To obtain the corresponding number of necessary SNe, one
can multiply the number of required fields in
Figures~\ref{fig:SNobs_45}b and \ref{fig:SNobs_35}b with the
corresponding number of SNe per unit redshift per field in
Figures~\ref{fig:SNobs_45}a and \ref{fig:SNobs_35}a, respectively.
Typically, the detection of the reionization drop with S/N $\gsim$ 3
requires at least $\langle N_{\rm pre}\rangle \gsim 10$ SNe.

The relative flatness of most of the curves in
Figures~\ref{fig:SNobs_45}b and \ref{fig:SNobs_35}b over the ranges $7
\lsim \zre \lsim 13$ and $6 \lsim \zre \lsim 10$, respectively, is due
to the fact that even though the number of detections decreases with
increasing redshift (making the detection of the reionization drop
more difficult), the separation between the various suppression curves
increases (making the detection of the reionization drop easier).  We
note from the figures that detecting the reionization drop in the SNe
distributions constructed from the 3.5 \micron~filter is more
efficient (requires fewer comparable fields) than using the 4.5
\micron~filter at redshifts 6 $\lsim$ $\zre$ $\lsim$ 10.  At redshifts larger
than $\zre\sim10$, the 4.5 \micron~filter becomes more efficient, and
can be used to detect the reionization drop out to $\zre\lsim15$ in a
1 year survey, in the most optimistic scenario.  However, such
conclusions about SNe at redshifts as high as $\zre\gsim$ 13--15 are
highly uncertain: (i) there could be an intrinsic physical cut--off to
the bright--end tail of the SNe luminosity distributions, drastically
reducing the number of detectable SNe at very high redshifts, and/or
(ii) detections in the 3.5 \micron~band, in addition to the the 4.5
\micron~band, could be necessary for photometric redshift
determination (see \S~\ref{sec:empirical}), which could further reduce
the number of very high redshift SNe with accurate redshift
determinations.

In the most optimistic scenario we considered (with the threshold for
star formation raised from $T_{\rm vir}(z>\zre) \gsim 300$ K to
$T_{\rm vir}(z<\zre) \gsim 1.1\times10^5$ K), corresponding to maximum
photo--heating feedback (abundant $\rm H_2$ cooling and no
self--shielding), the drop can be detected with just 8 (3.5 \micron)
fields over the range 6$\lsim$ $\zre$ $\lsim$10 and 20 (4.5 \micron) fields
over the range 6$\lsim\zre\lsim$13 with $\texp=10^5$s.  Using shorter
exposures, it can be detected with 20 (3.5 \micron) fields over the
range 6$\lsim\zre\lsim$10 and 100--300 (4.5 \micron) fields over the
range 6$\lsim\zre\lsim$13 with $\texp=10^4$s.  Less optimistic
scenarios can still allow for a detection of the reionization drop
with only tens of fields.  However, a reionization feature at $\zre
\gsim 15$, would require $\gsim 1000$ fields in order for the associated
drop to be detected.

In the most pessimistic scenario (with the threshold changing from
$T_{\rm vir} \gsim 10^4$ K $\rightarrow$ $10^4$ K), corresponding to
very strong self--shielding and no $\rm H_2$ at reionization (i.e. the
complete absence of any photo--heating feedback), reionization would not affect the
SNR and could not be detected.  Nevertheless, if a sharp reionization feature is
detected independently (through Ly$\alpha$ absorption spectra of
galaxies or quasars, 21-cm signatures, or CMB anisotropies), the lack
of a detected drop in the SNR would provide valuable evidence that
low--mass halos can withstand the effects of photo-ionization heating.  Note that without a complimenting detection of a sharp reionization feature, one can not distinguish between a very smooth, gradual reionization and an absence of photo--heating feedback.

Finally, we comment on the origin of the spike at $\zre\sim6$ in the
curves in Figure \ref{fig:SNobs_45}b. Apparently, the detection of the
reionization feature in the 4.5 \micron~band at $\zre\sim 6-7$ has an additional difficulty. This is caused by a combination of two effects: (i) the
decreasing separation between the observable SNe distributions (as
seen in Figure~\ref{fig:SNobs_45}a) at lower redshifts, and (ii) a
large--equivalent--width H$\alpha$ emission line in the spectrum of
our template SNe, SN1999em, at the rest frame wavelength $\lambda_{\rm
H \alpha} \approx 6563$ \AA. (The latter effect also produces the
slight bump in the curves in Figure~\ref{fig:SNobs_45}a at the
redshift in which the 4.5 \micron~filter probes the line, $\zline \sim
4.5 \rm \mu m / \lambda_{H\alpha} - 1 ~ \sim ~ 6$.)  The strong
H$\alpha$ emission line influences the $K$-correction so as to
increase the number of detectable SNe at $\zline$.  This reduces the
apparent drop in the SFR, if the reionization feature occurs at redshifts just
above $\zline$ (and, likewise, increases the size of the drop if the
reionization feature occurs just below $\zline$). The detectability of the
reionization feature is only affected at $\zline\sim6$ due to the
combination of (i) and (ii) above. The presence of the H$\alpha$ line
is nevertheless potentially important, because reionization could have
occurred around this redshift \citep{MH04}.  This highlights the need
for an improved statistical description of SNe template spectra from
low-redshift data, allowing one to ``model out'' an offending emission
line, or for observations of the high--$z$ SNe in a different filter
(as can be seen from Figure~\ref{fig:SNobs_35}, using the 3.5
\micron~filter bypasses this issue).

In order to obtain high--redshift SNRs, several simplifications have
been made in our analysis, involving the poorly constrained redshift
evolution (more precisely, the lack of evolution) of the stellar
initial mass function (IMF), SN progenitors, star formation
efficiency, SN lightcurves and spectra, etc.  Because of this, the
numbers presented in Figures \ref{fig:SNobs_45} and \ref{fig:SNobs_35}
should be regarded only as rough estimates of what a future survey
might achieve.  However, our main conclusions rely on a sudden feature
in the redshift evolution of the SNR, and should remain valid, as long
as evolution in other parameters occurs over a much wider redshift
range than the reionization feature \citep{WL04_size, BL04}, and as
long as the pre-reionization SNR is not significantly reduced (since
the latter would diminish the confidence at which any subsequent drop
in the SNR, caused by reionization, could be detected).

\subsubsection{Pop--III SNe}
\label{sec:popIII}

As discussed above, high-redshift SNe whose progenitor stars are
formed from metal--free gas within minihalos could be intrinsically
very different from the low-redshift SNe, due to differences in the
progenitor environments (e.g. very low metalicities; \citealt{ABN02,
BCL02}).  If such differences could be identified and detected, then
these ``pop--III'' SNe could provide valuable information about
primordial stars and their environments.  Indeed, \citet{WA05}
recently studied the redshift--distribution of such primordial SNe,
but only briefly addressed the issue of their detectability.  In order to
directly assess the number of such SNe among the hypothetical SNe
samples we obtained here, in Figure~\ref{fig:mini_f}, we plot the
fraction of SNe whose progenitor stars are located in minihalos.  The
solid curve assumes our fiducial model with a Salpeter IMF and
$\epsilon_{\ast \rm minihalo} = 0.1$; the dotted curve assumes that
each minihalo produces only a single star, and hence a single SN, over
a dynamical time (assuming that strong feedback from this star
disrupts any future star formation; as in \citealt{WA05}).  The
figure shows that with an unevolving star formation efficiency,
progenitor stars in minihalos would account for over half of the
SNe at $z\gsim9$.  On the other hand, in the extreme case of a single
SNe per minihalo, progenitor stars in minihalos would account for less
than $\sim 1\%$ of the $z\sim 10$ SNe (although at the earliest
epochs, $z\gsim22$, they would still constitute over half of all
SNe).  Given the overall detection rate of several hundred $z\gsim10$
SNe with a 1 year {\it JWST} survey we have found above, even in this
extreme case, Figure~\ref{fig:mini_f} implies that several of these
SNe could be caused by pop--III stars.

\vspace{+0\baselineskip}
\myputfigure{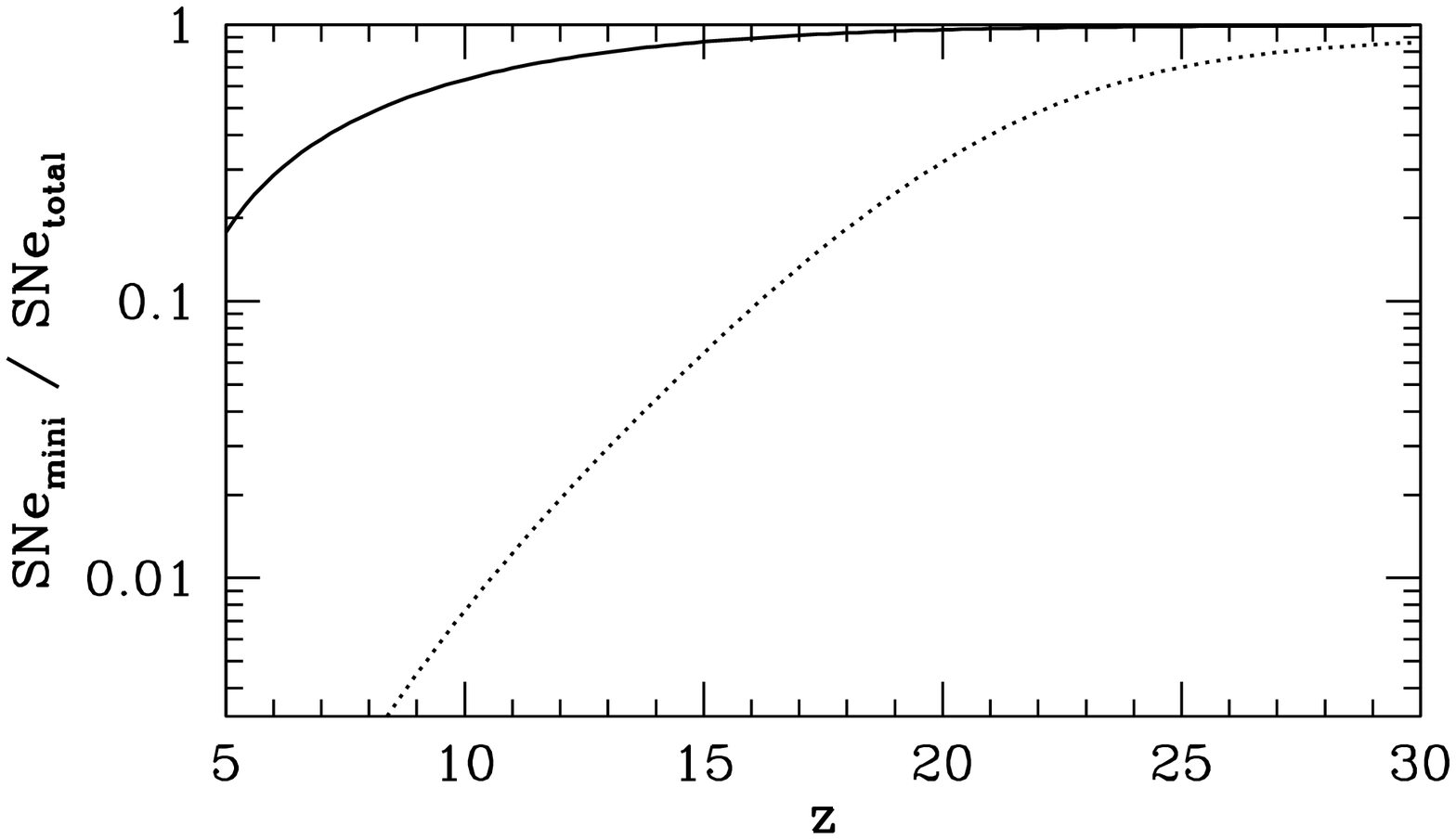}{3.3}{0.5}{.}{0.}  
\vspace{-1\baselineskip} \figcaption{Fraction of SNe whose progenitor
stars are located in minihalos.  The solid curve assumes our fiducial
model with a Salpeter IMF and $\epsilon_{\ast \rm minihalo} = 0.1$;
the dotted curve assumes that each minihalo produces only a single
star (and therefore a single SN) over a dynamical time.
\label{fig:mini_f}}
\vspace{+1\baselineskip}

\subsubsection{Varying Minihalo Star--Formation Efficiency}
\label{sec:eff}

The uncertainties in the typical star--formation efficiency in
minihalos can also directly influence our estimates for the
detectability of the reionization feature.  We have already shown that
the feature is detectable from SNe distributions even in the absence
of star formation in minihalos, but only provided that halos do not
self-shield effectively and the virial temperature for star--formation
is raised well above $10^4$K (e.g. see the right panels of figures
\ref{fig:SNobs_45}b and \ref{fig:SNobs_35}b).  If self-shielding is
very effective, the reionization drop is determined by the evolution of the
SFR in minihalos.  In Figure~\ref{fig:mini_eff}, we plot the S/N with
which the reionization feature is detectable from a fiducial
one--year survey, as a function of the assumed (constant)
star--formation efficiency in minihalos, $\epsilon_{\ast \rm
minihalo}$.  The three curves correspond to $\zre$ = 7 ({\it solid
curve}), 10 ({\it dotted curve}), 13 ({\it dashed curve}).  All curves
assume the transition $T_{\rm vir}(z>\zre) \gsim 300$ K $\rightarrow$ $T_{\rm vir}(z<\zre) \gsim 10^4$ K, as well as a flux density
threshold of 3 nJy ($\texp=10^5$s with the 4.5 \micron~{\it JWST}
filter), and no dust extinction.  The figure shows that in this case,
the reionization signal is detectable at $6\lsim\zre\lsim13$ with S/N
$>2$ as long as the star formation efficiency in minihalos exceeds
$\epsilon_{\ast \rm minihalo} \sim 0.05$.

\vspace{+0\baselineskip} \myputfigure{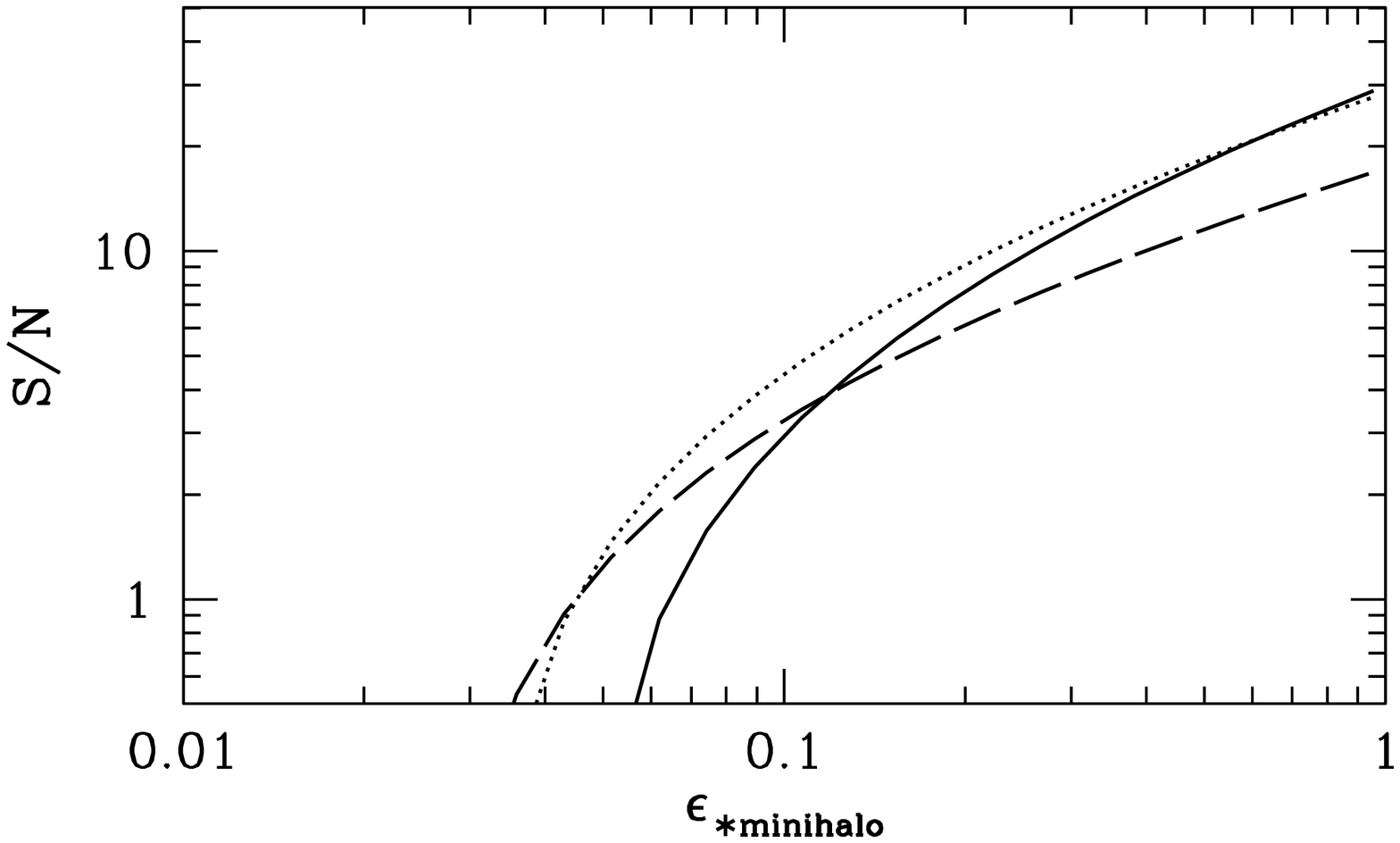}{3.3}{0.5}{.}{0.}
\vspace{-1\baselineskip} \figcaption{S/N with which reionization can
be detected as a function of the assumed (constant) star--formation
efficiency in minihalos, $\epsilon_{\ast \rm minihalo}$.  The three
curves correspond to three different reionization redshifts $\zre$ = 7
({\it solid curve}), 10 ({\it dotted curve}), and 13 ({\it dashed
curve}).  All curves assume the transition $T_{\rm vir}(z>\zre) \gsim 300$ K $\rightarrow$ $T_{\rm vir}(z<\zre) \gsim 10^4$ K, as well as a flux density threshold of 3 nJy
($\texp=10^5$s with the 4.5 \micron~{\it JWST} filter), no dust
extinction, and a $\tsurv = 1$yr survey (or, correspondingly, using
1yr/(2$\texp$) images with a $2.3'\times4.6'$ FOV).
\label{fig:mini_eff}}
\vspace{+1\baselineskip}

\subsubsection{Varying $\deltazre$}
\label{sec:vary_zre}

As mentioned previously, the width of a reionization feature, which we have
parameterized by $\deltazre$ (see the clarification below
eq.(\ref{eq:StoN})), is uncertain.  In our fiducial model we have
assumed to study fairly sharp features with $\deltazre=1$.  Here
we investigate the detectability of more extended
reionization features. In Figure~\ref{fig:delz}, we plot the maximum
reionization redshift, $z_{\rm re~max}$, up to which the reionization
signal would be detectable at S/N $\gsim$ 3 in a 1 yr survey, as a
function of $\deltazre$.  To approximate the redshift-smearing due to
any self-regulation or scatter, we simply follow the procedure
discussed above (following eq.\ref{eq:StoN}), but take averages of the
SNR in increasingly wider redshift bins prior to and following $\zre$.
This effectively averages the SNR in two bins whose mean redshifts are
separated by $\deltazre$ and mimics a suppression that is spread-out
over this redshift, rather than instantaneous. A reionization feature occurring
in the range $6 \lsim \zre \lsim z_{\rm re~max}$, smeared over a
redshift range of $\deltazre$, would be detectable out to the
redshifts shown in

\vspace{+0\baselineskip}
\myputfigure{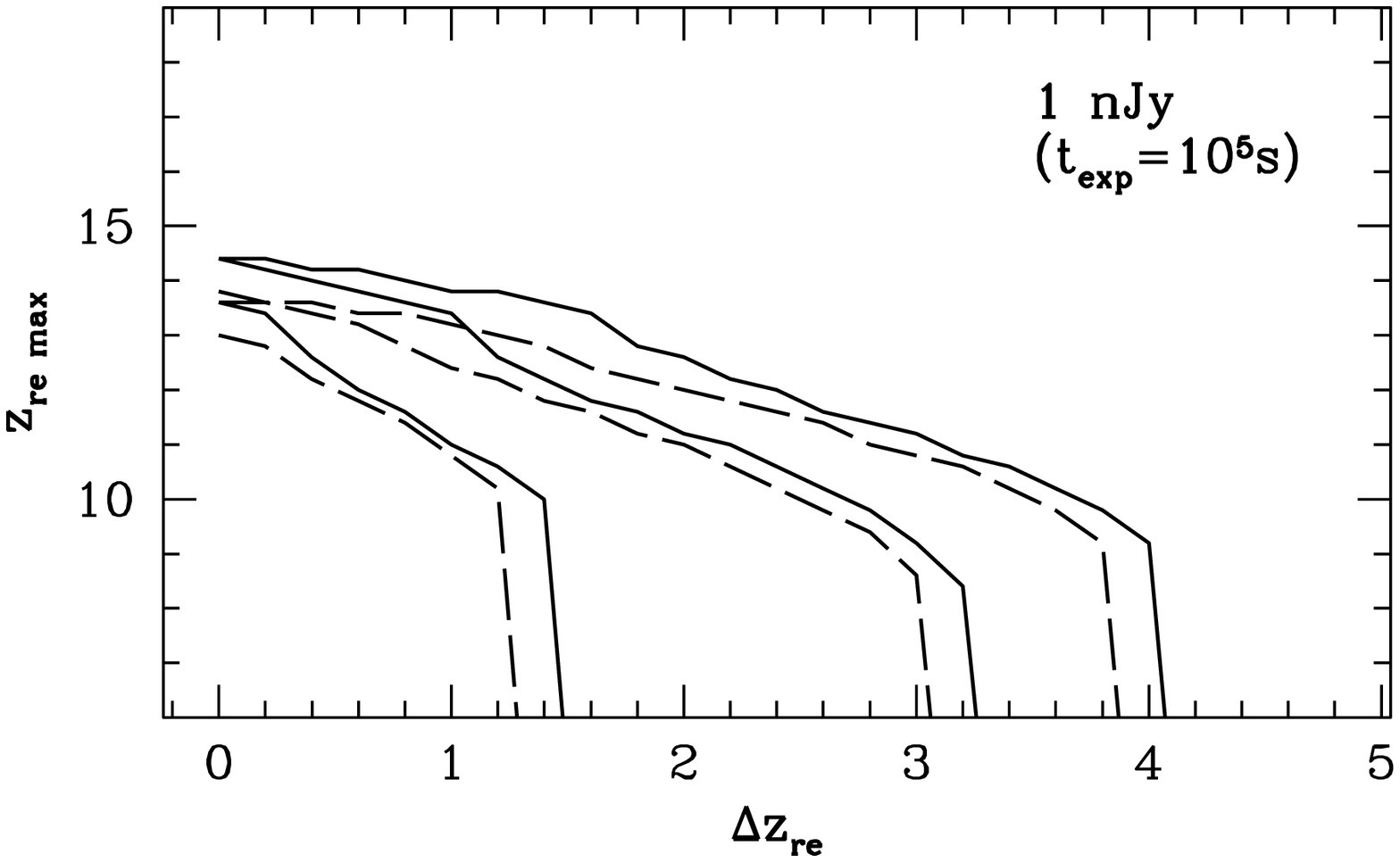}{3.3}{0.5}{.}{0.}  
\vspace{-1\baselineskip} \figcaption{Maximum reionization redshift,
$z_{\rm re~max}$, up to which the reionization signal would be
detectable at S/N $\gsim$ 3 in a 1 yr survey, as a function of
$\deltazre$.  The curves correspond to the reionization transitions of
$T_{\rm vir} \gsim 300$ K $\rightarrow$ $10^4$ K, $T_{\rm vir} \gsim
300$ K $\rightarrow$ $4.5\times10^4$ K, and $T_{\rm vir} \gsim 300$ K
$\rightarrow$ $1.1\times10^5$ K, ({\it left to right}).  Solid curves
correspond to no dust obscuration; dashed curves include dust.  All
curves assume a detection threshold of 1 nJy, analogous to
$\texp=10^5$ s exposures in the 3.5 \micron~{\it JWST} band.
\label{fig:delz}}
\vspace{+1\baselineskip}

\noindent
 Figure~\ref{fig:delz}.  The curves correspond to
the reionization transitions of $T_{\rm vir} \gsim 300$ K
$\rightarrow$ $10^4$ K, $T_{\rm vir} \gsim 300$ K $\rightarrow$
$4.5\times10^4$ K, and $T_{\rm vir} \gsim 300$ K $\rightarrow$
$1.1\times10^5$ K, ({\it left to right}).  Solid curves correspond to
no dust obscuration; dashed curves include dust.  All curves assume a
detection threshold of 1 nJy, analogous to $\texp=10^5$ s exposures in
the 3.5 \micron~{\it JWST} band.  From the figure, one can note that
the most optimistic scenario ($T_{\rm vir} \gsim 300$ K $\rightarrow$
$1.1\times10^5$ K), a reionization feature is detectable even if it occurs over
a redshift range $\deltazre \sim 4$.  In other cases, the detection of
the reionization feature requires $\deltazre \lsim$ 1 -- 3.

\subsubsection{Feasibility of a Fairly Sudden Drop in SNe Rates}
\label{sec:sharp}

The details and duration of the reionization epoch, and hence the shape and width of the drop in SNe rates, are unknown at this time.  Results from {\it WMAP} and the SDSS QSOs offer suggestive, albeit not conclusive, evidence that reionization was extended in redshift.\footnote{Note that there are two, often confused processes and time-scales associated with reionization: (i) the increase in intensity of the ionizing background and the mean free path of ionizing photons and (ii) the increase in the filling factor of ionized regions.  For the purposes of this paper, we concern ourselves with (ii).}  If reionization is indeed quite extended in redshift ($\Delta z \sim 10$), the associated drop in the SFR and SNR could be smeared out too much to be detectable.  However, even in the pessimistic (as it pertains to our analysis below) scenario where some effective {\it width} of the reionization feature is as large as $\Delta z \sim 10$, the {\it shape} of the feature need not be smooth, and could contain ``sharp'' ($\deltazre \sim 1$) drops.  Such transitions are probable in reionization histories in which different sources dominate different epochs (e.g. \citealt{HH03}).

The relevant process in determining the width and shape of the drop in SNRs is the evolution of the volume filling factor of ionized regions and their correlation with the small, vulnerable halos\footnote{Note the distinction here between ``small halos'' and ``minihalos''.  In the discussion below, we use the term small halos to denote all halos whose SFRs will be suppressed by reionization.  Thus small halos include minihalos as well as larger halos, depending on their susceptibility to negative photo-heating feedback.} whose SFRs are sensitive to the thermal state of the IGM.  In other words, the sharpness of the drop in SNRs, in each epoch during reionization, will depend on:
\begin{enumerate}
\item the nature of the dominant ionizing sources
\item the ionizing efficiency of the dominant ionizing sources
\item the level of synchronization of small halo formation with the formation of dominant ionizing sources
\item the level of synchronization of the dominant ionizing sources
\end{enumerate}

If the small, vulnerable halos themselves are the dominant ionizing sources at that particular epoch, then a sharp reionization feature could result if: (2) is high, and (3) is moderate (in this case (3) and (4) are the same).  If (3) is too high (i.e. the formation of small, vulnerable halos is very clustered in time and space), then negative feedback from the ionizing radiation could delay substantial growth of HII bubbles until small halos are no longer forming prodigiously enough to serve as signposts for reionization.  If (3) is too low, there might not be enough affected halos to notice the suppression; or large, isolated patches might have to wait a long time for their own ionizing sources to form (if (2) is not very high), thus smearing out the signal through pure cosmic variance.

If the small, vulnerable halos are not the dominant ionizing sources (as would be expected for the later periods of an extended reionization), then a sharp feature could result if (2) is high enough to reasonably counter cosmic variance.  However, we could relax our fine-tuning on (3) above, since feedback no-longer hinders the growth of HII regions.  There merely need to be enough small halos at that epoch to act as signposts for reionization.  From Figure \ref{fig:mini_f}, we see that this a reasonable assumption, especially given the fact that most small halos which are still forming at such late stages are probably not going to be very near the large overdensities which were likely to be ionized during earlier stages \citep{FO05, RGS02}.  We also require (4) to be reasonably high (i.e. that the dominant ionizing sources appear around the same time, without too much cosmic scatter).  Below, we further quantify such a scenario.

One can get a sense of the possible shapes of the reionization feature through an estimate of the evolution of the filling factor of ionized regions, $F_{\rm HII}(z)$, (c.f. \citealt{BL01, HH03}):
\begin{equation}
\label{eq:F_HII}
\frac{dF_{\rm HII}(z)}{dt} = \epsilon_\ast f_{\rm esc} \frac{N_{\rm ph/b}}{0.76} \frac{dF_{\rm col}(>M_{\rm min}(z), z)}{dt} - \alpha_B C \langle n^0_{\rm H} \rangle (1+z)^3 F_{\rm HII} ~ .
\end{equation}
Here $f_{\rm esc}$ is the escape fraction of ionizing photons, $N_{\rm ph/b}$ is the number of ionizing photons per baryon emitted by a typical source, $F_{\rm col}(>M, z)$ is the fraction of baryons that reside in collapsed halos with a total mass greater than $M$ at redshift z, $\alpha_B$ is the hydrogen case B recombination coefficient, $C \equiv \langle n_{\rm H}^2 \rangle / \langle n_{\rm H} \rangle^2$ is the clumping factor, and $\langle n^0_{\rm H} \rangle$ is the current hydrogen number density.  The first term on the right hand side accounts for ``new'' ionizations contributing to the growth of the HII regions and the last term on the right hand side accounts for ``old'' reionizations due to recombinations inside the HII region.  This equation is a very rough approximation, as it does not include feedback effects, light travel time, and it does not accurately model the period when bubbles start overlapping (i.e. $F_{\rm HII}(z) \sim 1$).  However, it can suffice for the crude estimates we are making here.

In Figure \ref{fig:F_HII}, we plot $F_{\rm HII}(z)$ for several values of $M_{\rm min}(z)$ corresponding to redshift--independent values of $\Tvir =$ 300 K, $10^4$ K, and $10^5$ K, from right to left in the figure.  The plot assumes values of ($\epsilon_\ast$, $f_{\rm esc}$, $N_{\rm ph/b}$, $C$) = (0.1, 0.1, 4000, 4).  From the figure, it is evident that fairly rapid growth of $F_{\rm HII}(z)$ is possible for $F_{\rm HII}(z) \gsim 0.1$.  For example, the $\Tvir \gsim 10^5$ K curve goes from $F_{\rm HII} \sim 0.3$ to $F_{\rm HII} \sim 0.8$ in a the redshift interval $z \sim$ 8.5 $\rightarrow$ 7.  So SNe in roughly 50\% of small halos could be wiped out in a redshift interval of only $\Delta \zre \sim 1.5$.  Comparing with Figures \ref{fig:SNobs_45}a, \ref{fig:SNobs_35}a or \ref{fig:mini_f}, one can see that this could result in a fairly large, easily detectable drop in the SNRs.

To summarize, we outline a likely reionization scenario.  An early reionization epoch could be driven by minihalos.  Feedback processes can stall reionization at a constant or even decreasing filling factor of ionized regions.  Then a later population of more massive halos ($\Tvir \sim$ $10^4$ -- $10^5$ K) could complete the ionization process on time-scales corresponding to $\deltazre \sim$ 1 -- 2.  The research we present here suggests that this later epoch is detectable through an accompanying drop in the SNRs.

\vspace{+0\baselineskip}
\myputfigure{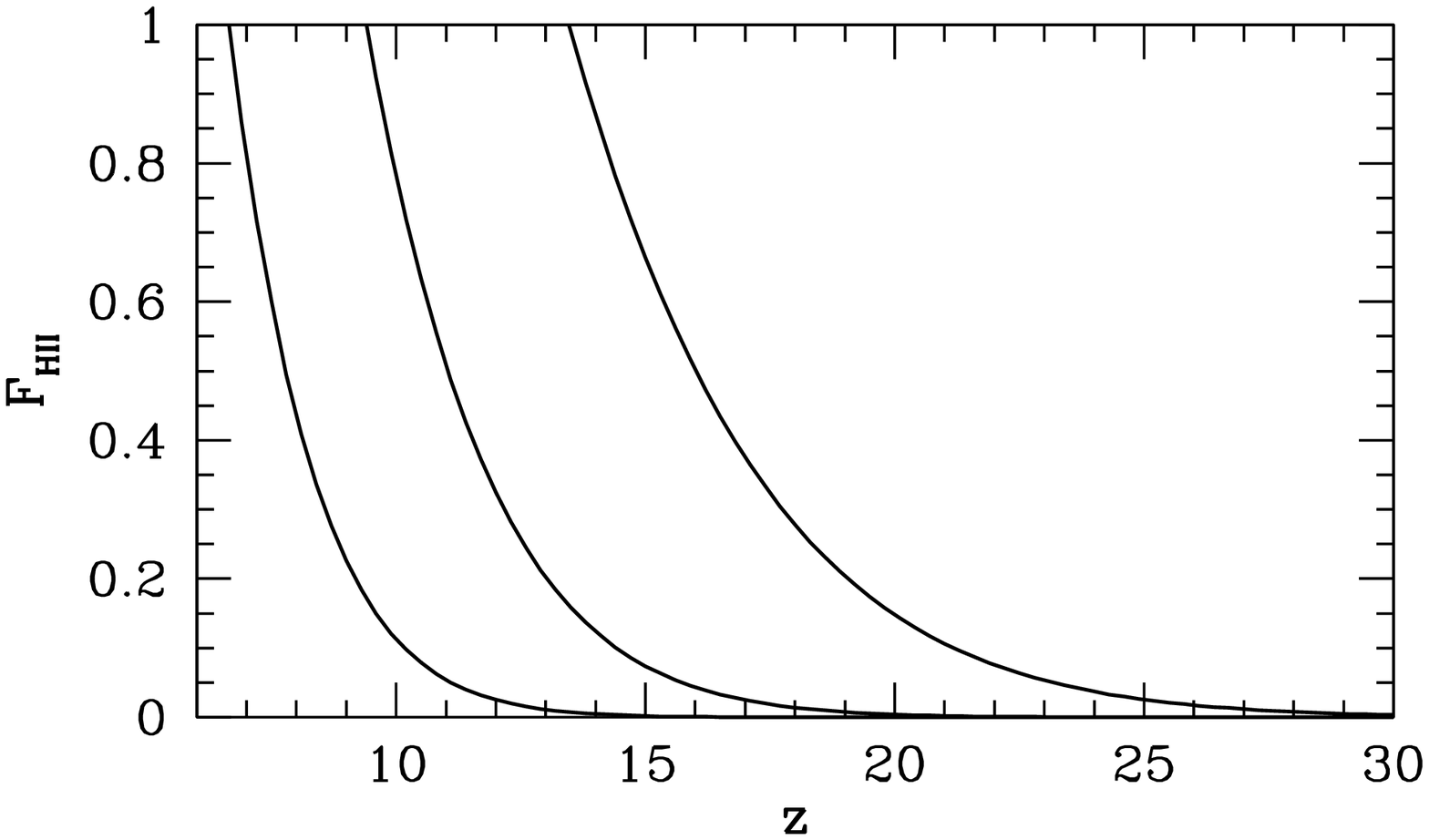}{3.3}{0.5}{.}{0.}  
\vspace{-1\baselineskip} \figcaption{
Simple models of the evolution of the filling factor of ionized regions, $F_{\rm HII}(z)$, for several values of $M_{\rm min}(z)$ corresponding to redshift--independent values of $\Tvir =$ 300 K, $10^4$ K, and $10^5$ K, {\it right to left}.
\label{fig:F_HII}}
\vspace{+1\baselineskip}

\section{Conclusions}
\label{sec:conclusions}

We obtain detailed, high-redshift SNe rates which could be detected in
future infrared surveys, such as {\it JWST}.  Our method uses the
extended Press-Schechter formalism to generate high-redshift SFRs and
SNRs. We empirically calibrate our model using the observed properties
of local SNe, such as their lightcurves, fiducial spectra, as well as
dust corrected and uncorrected peak magnitude distributions, to
predict the detectable high--$z$ SNRs.

During reionization, the ionizing background radiation heats the IGM
and could dramatically suppress gas accretion and star formation in
low--mass halos.  As a result, reionization may be accompanied by a
drop in the SFR and the corresponding SNR.  The size of the drop is
uncertain, since the ability of the gas in low--mass halos to cool and
self--shield against the ionizing radiation is poorly constrained at
high redshifts.  However, we argued, using specific illustrative
models, that a relatively sharp feature is feasible. Using the SNRs we
generate, we then analyze the prospects of detecting such a feature with
future SNe observations, specifically those obtainable with {\it
JWST}.

We find that 4 -- 24 SNe may be detectable from $z\gsim5$ at the
sensitivity of 3 nJy (requiring $10^5$ s exposures in a 4.5
\micron~band) in each $\sim 10$ arcmin$^2$ {\it JWST} field.  In a
hypothetical one year survey, we expect to detect up to thousands of
SNe per unit redshift at $z\sim6$.  Our results imply that, for most
scenarios, high-redshift SNe observations can be used to detect fairly
sharp features in the reionization history (occurring over $\deltazre
\sim$ 1 -- 3) out to $z \sim 13$, as well as set constraints on the
photo-ionization heating feedback on low--mass halos at the
reionization epoch.  Specifically, for a wide range of scenarios at
$\zre \lsim 13$, the drop in the SNR due to reionization can be
detected at S/N $\gsim$ 3 with only tens of deep {\it JWST} exposures.
Our results therefore suggest that future searches for high--$z$ SNe
could be a valuable new tool, complementing other techniques, to study
the process of reionization, as well as the feedback mechanism that
regulates it.

\acknowledgments{We thank Rosalba Perna, David Helfand, Avi Loeb, Tom
Abel, Rennan Barkana, Peng Oh, and Steven Furlanetto for useful
discussions.  ZH gratefully acknowledges support by the NSF through
grants AST-0307200 and AST-0307291, and by NASA through grants
NNG04GI88G and NNG05GF14G.}

\vspace{-2\baselineskip}

\bibliographystyle{apj}
\bibliography{apj-jour,ms}

\begin{thebibliography}{60}
\expandafter\ifx\csname natexlab\endcsname\relax\def\natexlab#1{#1}\fi

\bibitem[{{Abel} {et~al.}(2002){Abel}, {Bryan}, \& {Norman}}]{ABN02}
{Abel}, T., {Bryan}, G.~L., \& {Norman}, M.~L. 2002, Science, 295, 93

\bibitem[{{Adelberger} \& {Steidel}(2000)}]{AS00}
{Adelberger}, K.~L., \& {Steidel}, C.~C. 2000, \apj, 544, 218

\bibitem[{{Barkana} \& {Loeb}(2000)}]{BL00}
{Barkana}, R., \& {Loeb}, A. 2000, \apj, 539, 20

\bibitem[{{Barkana} \& {Loeb}(2001)}]{BL01}
---. 2001, \physrep, 349, 125

\bibitem[{{Barkana} \& {Loeb}(2004{\natexlab{a}})}]{BL04_grbvsqso}
---. 2004{\natexlab{a}}, \apj, 601, 64

\bibitem[{{Barkana} \& {Loeb}(2004{\natexlab{b}})}]{BL04}
---. 2004{\natexlab{b}}, \apj, 609, 474

\bibitem[{{Baron} {et~al.}(2000){Baron}, {Branch}, {Hauschildt}, {Filippenko},
  {Kirshner}, {Challis}, {Jha}, {Chevalier}, {Fransson}, {Lundqvist},
  {Garnavich}, {Leibundgut}, {McCray}, {Michael}, {Panagia}, {Phillips}, {Pun},
  {Schmidt}, {Sonneborn}, {Suntzeff}, {Wang}, \& {Wheeler}}]{Baron00}
{Baron}, E. {et~al.} 2000, \apj, 545, 444

\bibitem[{{Bromm} {et~al.}(2002){Bromm}, {Coppi}, \& {Larson}}]{BCL02}
{Bromm}, V., {Coppi}, P.~S., \& {Larson}, R.~B. 2002, \apj, 564, 23

\bibitem[{{Bromm} {et~al.}(2001){Bromm}, {Ferrara}, {Coppi}, \&
  {Larson}}]{BFCL01}
{Bromm}, V., {Ferrara}, A., {Coppi}, P.~S., \& {Larson}, R.~B. 2001, \mnras,
  328, 969

\bibitem[{{Bromm} \& {Loeb}(2002)}]{BL02}
{Bromm}, V., \& {Loeb}, A. 2002, \apj, 575, 111

\bibitem[{{Bryan} \& {Norman}(1998)}]{BN98}
{Bryan}, G.~L., \& {Norman}, M.~L. 1998, \apj, 495, 80

\bibitem[{{Bunker} {et~al.}(2004){Bunker}, {Stanway}, {Ellis}, \&
  {McMahon}}]{Bunker04}
{Bunker}, A.~J., {Stanway}, E.~R., {Ellis}, R.~S., \& {McMahon}, R.~G. 2004,
  \mnras, 355, 374

\bibitem[{{Cappellaro} {et~al.}(1997){Cappellaro}, {Turatto}, {Tsvetkov},
  {Bartunov}, {Pollas}, {Evans}, \& {Hamuy}}]{C97}
{Cappellaro}, E., {Turatto}, M., {Tsvetkov}, D.~Y., {Bartunov}, O.~S.,
  {Pollas}, C., {Evans}, R., \& {Hamuy}, M. 1997, \aap, 322, 431

\bibitem[{{Cen}(2003)}]{Cen03}
{Cen}, R. 2003, \apjl, 591, L5

\bibitem[{{Cen} \& {McDonald}(2002)}]{CM02}
{Cen}, R., \& {McDonald}, P. 2002, \apj, 570, 457

\bibitem[{{Cen} \& {Ostriker}(1992)}]{CO92}
{Cen}, R., \& {Ostriker}, J.~P. 1992, \apjl, 399, L113

\bibitem[{{Choudhury} \& {Srianand}(2002)}]{CS02}
{Choudhury}, T.~R., \& {Srianand}, R. 2002, \mnras, 336, L27

\bibitem[{{Dahl{\' e}n} \& {Fransson}(1999)}]{DF99}
{Dahl{\' e}n}, T., \& {Fransson}, C. 1999, \aap, 350, 349

\bibitem[{{Dijkstra} {et~al.}(2004){Dijkstra}, {Haiman}, {Rees}, \&
  {Weinberg}}]{Dijkstra04}
{Dijkstra}, M., {Haiman}, Z., {Rees}, M.~J., \& {Weinberg}, D.~H. 2004, \apj,
  601, 666

\bibitem[{{Doggett} \& {Branch}(1985)}]{DB85}
{Doggett}, J.~B., \& {Branch}, D. 1985, \aj, 90, 2303

\bibitem[{{Efstathiou}(1992)}]{Efstathiou92}
{Efstathiou}, G. 1992, \mnras, 256, 43P

\bibitem[{{Fan} {et~al.}(2004){Fan}, {Hennawi}, {Richards}, {Strauss},
  {Schneider}, {Donley}, {Young}, {Annis}, {Lin}, {Lampeitl}, {Lupton}, {Gunn},
  {Knapp}, {Brandt}, {Anderson}, {Bahcall}, {Brinkmann}, {Brunner}, {Fukugita},
  {Szalay}, {Szokoly}, \& {York}}]{Fan04}
{Fan}, X. {et~al.} 2004, \aj, 128, 515

\bibitem[{Furlanetto \& Oh(2005)}]{FO05}
Furlanetto, S.~R., \& Oh, S.~P. 2005, astro-ph/0505065

\bibitem[{{Gabasch} {et~al.}(2004){Gabasch}, {Salvato}, {Saglia}, {Bender},
  {Hopp}, {Seitz}, {Feulner}, {Pannella}, {Drory}, {Schirmer}, \&
  {Erben}}]{Gabasch04}
{Gabasch}, A. {et~al.} 2004, \apjl, 616, L83

\bibitem[{{Giavalisco} {et~al.}(2004){Giavalisco}, {Dickinson}, {Ferguson},
  {Ravindranath}, {Kretchmer}, {Moustakas}, {Madau}, {Fall}, {Gardner},
  {Livio}, {Papovich}, {Renzini}, {Spinrad}, {Stern}, \&
  {Riess}}]{Giavalisco04}
{Giavalisco}, M. {et~al.} 2004, \apjl, 600, L103

\bibitem[{{Gnedin}(1996)}]{Gnedin96}
{Gnedin}, N.~Y. 1996, \apj, 456, 1

\bibitem[{{Gnedin}(2000)}]{Gnedin00b}
---. 2000, \apj, 542, 535

\bibitem[{{Haiman} {et~al.}(2000){Haiman}, {Abel}, \& {Rees}}]{HAR00}
{Haiman}, Z., {Abel}, T., \& {Rees}, M.~J. 2000, \apj, 534, 11

\bibitem[{{Haiman} \& {Holder}(2003)}]{HH03}
{Haiman}, Z., \& {Holder}, G.~P. 2003, \apj, 595, 1

\bibitem[{{Haiman} \& {Loeb}(1999)}]{HL99}
{Haiman}, Z., \& {Loeb}, A. 1999, \apj, 519, 479

\bibitem[{{Hamuy} {et~al.}(2001){Hamuy}, {Pinto}, {Maza}, {Suntzeff},
  {Phillips}, {Eastman}, {Smith}, {Corbally}, {Burstein}, {Li}, {Ivanov},
  {Moro-Martin}, {Strolger}, {de Souza}, {dos Anjos}, {Green}, {Pickering},
  {Gonz{\' a}lez}, {Antezana}, {Wischnjewsky}, {Galaz}, {Roth}, {Persson}, \&
  {Schommer}}]{H01}
{Hamuy}, M. {et~al.} 2001, \apj, 558, 615

\bibitem[{{Heger} {et~al.}(2003){Heger}, {Fryer}, {Woosley}, {Langer}, \&
  {Hartmann}}]{Heger03}
{Heger}, A., {Fryer}, C.~L., {Woosley}, S.~E., {Langer}, N., \& {Hartmann},
  D.~H. 2003, \apj, 591, 288

\bibitem[{{Holder} {et~al.}(2001){Holder}, {Haiman}, \& {Mohr}}]{HHM01}
{Holder}, G., {Haiman}, Z., \& {Mohr}, J.~J. 2001, \apjl, 560, L111

\bibitem[{{Kitayama} \& {Ikeuchi}(2000)}]{KI00}
{Kitayama}, T., \& {Ikeuchi}, S. 2000, \apj, 529, 615

\bibitem[{{Kogut} {et~al.}(2003){Kogut}, {Spergel}, {Barnes}, {Bennett},
  {Halpern}, {Hinshaw}, {Jarosik}, {Limon}, {Meyer}, {Page}, {Tucker},
  {Wollack}, \& {Wright}}]{Kogut03}
{Kogut}, A. {et~al.} 2003, \apjs, 148, 161

\bibitem[{{Lacey} \& {Cole}(1993)}]{LC93}
{Lacey}, C., \& {Cole}, S. 1993, \mnras, 262, 627

\bibitem[{{Leibundgut} \& {Suntzeff}(2003)}]{LS03}
{Leibundgut}, B., \& {Suntzeff}, N.~B. 2003, LNP Vol.~598: Supernovae and
  Gamma-Ray Bursters, 598, 77

\bibitem[{{Lilly} {et~al.}(1996){Lilly}, {Le Fevre}, {Hammer}, \&
  {Crampton}}]{Lilly96}
{Lilly}, S.~J., {Le Fevre}, O., {Hammer}, F., \& {Crampton}, D. 1996, \apjl,
  460, L1+

\bibitem[{{Madau} {et~al.}(1998){Madau}, {della Valle}, \& {Panagia}}]{MVP98}
{Madau}, P., {della Valle}, M., \& {Panagia}, N. 1998, \mnras, 297, L17+

\bibitem[{{Madau} {et~al.}(1996){Madau}, {Ferguson}, {Dickinson}, {Giavalisco},
  {Steidel}, \& {Fruchter}}]{Madau96}
{Madau}, P., {Ferguson}, H.~C., {Dickinson}, M.~E., {Giavalisco}, M.,
  {Steidel}, C.~C., \& {Fruchter}, A. 1996, \mnras, 283, 1388

\bibitem[{{Madau} {et~al.}(1997){Madau}, {Meiksin}, \& {Rees}}]{MMR97}
{Madau}, P., {Meiksin}, A., \& {Rees}, M.~J. 1997, \apj, 475, 429

\bibitem[{{Mesinger} \& {Haiman}(2004)}]{MH04}
{Mesinger}, A., \& {Haiman}, Z. 2004, \apjl, 611, L69

\bibitem[{{Mesinger} {et~al.}(2004){Mesinger}, {Haiman}, \& {Cen}}]{MHC04}
{Mesinger}, A., {Haiman}, Z., \& {Cen}, R. 2004, \apj, 613, 23

\bibitem[{{Mesinger} {et~al.}(2005){Mesinger}, {Perna}, \& {Haiman}}]{MPH05}
{Mesinger}, A., {Perna}, R., \& {Haiman}, Z. 2005, \apj, 623, 1

\bibitem[{{Miralda-Escude} \& {Rees}(1997)}]{MeR97}
{Miralda-Escude}, J., \& {Rees}, M.~J. 1997, \apjl, 478, L57+

\bibitem[{{Nemiroff}(2003)}]{N03}
{Nemiroff}, R.~J. 2003, \aj, 125, 2740

\bibitem[{{Patat} {et~al.}(1994){Patat}, {Barbon}, {Cappellaro}, \&
  {Turatto}}]{P94}
{Patat}, F., {Barbon}, R., {Cappellaro}, E., \& {Turatto}, M. 1994, \aap, 282,
  731

\bibitem[{{Richardson} {et~al.}(2002){Richardson}, {Branch}, {Casebeer},
  {Millard}, {Thomas}, \& {Baron}}]{Richardson02}
{Richardson}, D., {Branch}, D., {Casebeer}, D., {Millard}, J., {Thomas}, R.~C.,
  \& {Baron}, E. 2002, \aj, 123, 745

\bibitem[{{Ricotti} {et~al.}(2002){Ricotti}, {Gnedin}, \& {Shull}}]{RGS02}
{Ricotti}, M., {Gnedin}, N.~Y., \& {Shull}, J.~M. 2002, \apj, 575, 49

\bibitem[{{Shapiro} {et~al.}(1994){Shapiro}, {Giroux}, \& {Babul}}]{SGB94}
{Shapiro}, P.~R., {Giroux}, M.~L., \& {Babul}, A. 1994, \apj, 427, 25

\bibitem[{{Shapiro} {et~al.}(2004){Shapiro}, {Iliev}, \& {Raga}}]{SIR04}
{Shapiro}, P.~R., {Iliev}, I.~T., \& {Raga}, A.~C. 2004, \mnras, 348, 753

\bibitem[{{Sokasian} {et~al.}(2004){Sokasian}, {Yoshida}, {Abel}, {Hernquist},
  \& {Springel}}]{SYAHS04}
{Sokasian}, A., {Yoshida}, N., {Abel}, T., {Hernquist}, L., \& {Springel}, V.
  2004, \mnras, 350, 47

\bibitem[{{Somerville} {et~al.}(2001){Somerville}, {Primack}, \&
  {Faber}}]{SPF01}
{Somerville}, R.~S., {Primack}, J.~R., \& {Faber}, S.~M. 2001, \mnras, 320, 504

\bibitem[{{Spergel} {et~al.}(2003){Spergel}, {Verde}, {Peiris}, {Komatsu},
  {Nolta}, {Bennett}, {Halpern}, {Hinshaw}, {Jarosik}, {Kogut}, {Limon},
  {Meyer}, {Page}, {Tucker}, {Weiland}, {Wollack}, \& {Wright}}]{Spergel03}
{Spergel}, D.~N. {et~al.} 2003, \apjs, 148, 175

\bibitem[{{Strolger} {et~al.}(2004){Strolger}, {Riess}, {Dahlen}, {Livio},
  {Panagia}, {Challis}, {Tonry}, {Filippenko}, {Chornock}, {Ferguson},
  {Koekemoer}, {Mobasher}, {Dickinson}, {Giavalisco}, {Casertano}, {Hook},
  {Blondin}, {Leibundgut}, {Nonino}, {Rosati}, {Spinrad}, {Steidel}, {Stern},
  {Garnavich}, {Matheson}, {Grogin}, {Hornschemeier}, {Kretchmer}, {Laidler},
  {Lee}, {Lucas}, {de Mello}, {Moustakas}, {Ravindranath}, {Richardson}, \&
  {Taylor}}]{Strolger04}
{Strolger}, L. {et~al.} 2004, \apj, 613, 200

\bibitem[{{Thoul} \& {Weinberg}(1996)}]{TW96}
{Thoul}, A.~A., \& {Weinberg}, D.~H. 1996, \apj, 465, 608

\bibitem[{{Wang} {et~al.}(2004){Wang}, {Crotts}, {Garnavich}, {Priedhorsky},
  {Habib}, {Heitmann}, {Kutyrev}, {Moseley}, {Squires}, {Tegmark}, {Wright}, \&
  {JEDI}}]{Wang04}
{Wang}, Y. {et~al.} 2004, American Astronomical Society Meeting Abstracts, 205,

\bibitem[{Wise \& Abel(2005)}]{WA05}
Wise, J.~H., \& Abel, T. 2005, ApJ, submitted, astro-ph/0411558

\bibitem[{{Wyithe} \& {Loeb}(2004{\natexlab{a}})}]{WL04_size}
{Wyithe}, J.~S.~B., \& {Loeb}, A. 2004{\natexlab{a}}, \nat, 432, 194

\bibitem[{{Wyithe} \& {Loeb}(2004{\natexlab{b}})}]{WL04_nf}
---. 2004{\natexlab{b}}, \nat, 427, 815

\end{thebibliography}

\end{document}